\documentclass[prl,reprint, twocolumn,showpacs,preprintnumbers,amsmath,amssymb,nofootinbib,floatfix]{revtex4-1} 
\usepackage{graphicx}

\usepackage{mathrsfs}
\usepackage{hyperref}

\usepackage{slashed}

\usepackage{color}

\usepackage{gensymb}

\usepackage{amsmath}

\usepackage{amssymb,graphicx}
\newcommand{\bbGamma}{{\mathpalette\makebbGamma\relax}}
\newcommand{\makebbGamma}[2]{%
\raisebox{\depth}{\scalebox{1}[-1]{$\mathsurround=0pt#1\mathbb{L}$}}%
}

\allowdisplaybreaks[4]

\def \matrix #1 {\left(\begin{array}{cc} #1 \end{array}\right)}

\def\II{\hbox{{1}\kern-.25em\hbox{l}}}

\begin{document}

\title{Two-Loop Renormalization-Group Evolution for the Nucleon Distribution Amplitude}

\author{Yong-Kang Huang$^{a}$}
\email{huangyongkang@mail.nankai.edu.cn}

\author{Yao Ji$^{b}$}
\email{corresponding author: yaoji@cuhk.edu.cn}


\author{Bo-Xuan Shi$^{a}$}
\email{corresponding author: shibx@mail.nankai.edu.cn}

\author{Yu-Ming Wang$^{a}$}
\email{corresponding author: wangyuming@nankai.edu.cn}

\affiliation{
\vspace{0.2 cm}
${}^a$ School of Physics, Nankai University,
Weijin Road 94, Tianjin 300071, P.R. China \\
${}^b$ School of Science and Engineering, The Chinese University of Hong Kong,
Shenzhen, Guangdong, 518172,  China
\vspace{0.2 cm}
}

\date{\today}

\begin{abstract}
\noindent
We determine for the first time the two-loop renormalization-group (RG) equation  for the nucleon light-cone distribution amplitude,
which constitutes the last missing ingredient for  the complete  next-to-leading-logarithmic corrections
to the nucleon form factors in the hard-collinear factorization framework.
Applying the conformal expansion for this fundamental nucleon distribution amplitude then enables us to construct
an  analytic solution that  captures the desired  scale dependence of phenomenologically interesting  series coefficients.
Importantly, the two-loop  RG evolutions  of these central hadronic  quantities can bring about  noticeable impacts
on the corresponding leading-logarithmic results   for  three sample models of the nucleon distribution amplitude.
\\[0.4em]

\end{abstract}


\maketitle

%
\section{Introduction}
%

Hadron distribution amplitudes on the light-cone defined by non-forward  matrix elements  of composite QCD  operators
are of fundamental importance for the systematic description of hard  exclusive  reactions in the field-theoretic framework.
In particular, they  open up new avenues for probing the intricate hadron structure properties in terms of quark/gluon degrees of freedom
when compared with the conventional parton distribution functions.
The  light-cone distribution amplitudes (LCDAs) of the nucleon in QCD  further  serve  as  an indispensable ingredient
in  the hard-collinear factorization formula for the nucleon electromagnetic form factor \cite{Lepage:1979za,Chernyak:1984bm,Huang:2024ugd,Chen:2024fhj},
which  belongs to the simplest and  most significant  observables of hadron physics.
These non-perturbative functions  also encode   the collinear dynamics of the  vacuum-to-nucleon correlation function
suitable for the construction of the light-cone sum rules of  the semileptonic $\Lambda_b \to p \, \ell \, \bar \nu_{\ell}$
form factors \cite{Khodjamirian:2011jp,Khodjamirian:2023wol},
which are  highly beneficial for  unraveling  the ultimate nature of the long-standing  discrepancy between
the exclusive and inclusive determinations of  the Cabibbo-Kobayashi-Maskawa (CKM) matrix element $|V_{ub}|$ \cite{ParticleDataGroup:2024cfk,HeavyFlavorAveragingGroupHFLAV:2024ctg}.
Consequently, advancing our understanding towards both the perturbative features \cite{Braun:1999te,Braun:2008ia,Braun:2009vc,Ji:2014eta}
and  the non-perturbative behaviours \cite{Chernyak:1984bm,King:1986wi,Chernyak:1987nv,QCDSF:2008qtn,Braun:2014wpa,Bali:2015ykx,RQCD:2019hps,Bali:2024oxg}
of the nucleon distribution amplitudes, at leading twist and beyond
(see \cite{Braun:2000kw,Braun:2006hz,Anikin:2013aka} for a general classification),
 has therefore triggered  intense theoretical activities over the past decades,
in order to further  improve   our theory predictions for these flagship hadron form factors.

Needless to say,  an in-depth exploration  of  perturbative  properties  of  the  nucleon   LCDAs  necessitates
controlling the renormalization-scale dependence of  such collinear functions.
The yielding renormalization-group (RG)  evolution equations are  in demand for establishing  QCD factorization formulae
of numerous baryon-to-nucleon transition matrix elements and for accomplishing  an all-order summation of
the parametrically large logarithms entering  the short-distance coefficient functions.
Uncovering  the  underlying   mathematical structure of these integro-differential equations
has therefore attracted enormous interest \cite{Braun:2013tva,Braun:2016qlg}
(see  \cite{Braun:2003rp,Belitsky:2004cz}  for an excellent review),
thanks to its intimate connection with conformal symmetry of the three-particle quantum-mechanical system.
In contrast with the celebrated Efremov-Radyushkin-Brodsky-Lepage (ERBL) evolution for the  two-particle meson distribution amplitude \cite{Lepage:1979zb,Efremov:1979qk}, the conformal symmetry does not even allow for constructing an exact solution to the  RG  equation
of the leading-twist nucleon distribution amplitude $\Phi_{N}$ \cite{Braun:1999te}.
Remarkably, the two lowest anomalous dimensions for the twist-three  nucleon distribution amplitude turn out to
decouple from the remaining  spectrum by a finite ``mass gap" $\Delta  = - (0.32 \pm 0.02)$,
which further   manifests in the obtained   spectrum of   anomalous dimensions for
the three-particle $B$-meson  distribution amplitude $\Phi_3$ \cite{Braun:2015pha}
and for the leading-twist $\Lambda_b$-baryon distribution amplitude  $\Phi_{\Lambda_b}$ \cite{Braun:2014npa}
in heavy quark effective theory.
The very  structure of the energy spectrum for the RG evolution kernel of the three-particle light-ray operator discussed above
is undoubtedly of decisive importance for  the model-independent extraction of  the asymptotic behaviour of $\Phi_{N}$
in the formal $\mu \to \infty$ limit, which is in turn  crucial to ensure  the convergence of the convolution integral
in the factorized expression of the  nucleon electromagnetic form factor.

As a matter of fact,  the lowest-order (one-loop) computation of the evolution kernel for
the leading-twist nucleon distribution amplitude had been already carried out in  \cite{Lepage:1979za,Lepage:1980fj},
with the standard diagrammatic approach,   more than  forty years ago.
However, the next-to-leading-order (NLO) QCD correction to this  three-particle RG kernel remains elusive even now,
due to the apparent technical challenges of implementing the two-loop ultraviolet (UV) renormalization for
the non-local baryonic operator  in the presence of evanescent operators \cite{Breitenlohner:1977hr,Bonneau:1980zp,Bonneau:1979jx,Collins:1984xc}
(see \cite{Dugan:1990df,Herrlich:1994kh,Buras:1989xd,Wang:2017ijn,Gao:2019lta,Li:2020rcg,Gao:2021iqq,Cui:2023jiw,Ji:2024iak,Ji:2018yaf}  for  additional discussions in a variety of contexts).
By contrast, the two-loop evolution kernel of the leading-twist pion distribution amplitude $\Phi_{\pi}$ became available
in the middle 1980s \cite{Sarmadi:1982yg,Dittes:1983dy,Katz:1984gf,Mikhailov:1984ii,Belitsky:1999gu}
and even the three-loop QCD computation of that two-particle ERBL kernel was recently pursued
with an attractive technique based upon conformal asymmetry \cite{Braun:2017cih,Ji:2023eni}.
Moreover,  the complete next-to-leading-logarithmic (NLL)  calculation of the Dirac nucleon form factor
in the perturbative factorization formalism cannot be  achieved  without determining   the RG  evolution  of the nucleon distribution amplitude  at the two-loop accuracy.
It is our primary objective to fill such an important and long-overdue gap in this Letter,
by establishing   the  desired NLO evolution equation  with the modern effective field theory approach
and  by constructing an explicit solution to this integro-differential equation with the conformal wave expansion of $\Phi_{N}$.
We then proceed to  derive  analytically the two-loop matching relation of the leading-twist nucleon   distribution amplitude
between our renormalization prescription  and  the Krankl-Manashov (KM)  scheme \cite{Krankl:2011gch},
which  constitutes  an essential ingredient of the perturbative  demonstration for the factorization-scheme independence
of  the nucleon electromagnetic  form factor.
Phenomenological significance of the  newly determined two-loop RG evolution of the twist-three nucleon distribution amplitude
will be explored subsequently with three sample models for $\Phi_{N}$.

%
\section{General Analysis}
%

The leading twist-three  nucleon distribution amplitude can be defined conveniently
in terms of  the renormalized three-particle light-ray operator  matrix element \cite{Braun:2000kw}
\begin{widetext}
\begin{eqnarray}
&& \left \langle 0 \left |   \epsilon_{i j k} \, \left [  u_{i^{\prime}}^{\uparrow}(\tau_1 \, n)[\tau_1 \, n,  \, \tau_0 \, n]_{i^{\prime} i} \,\, C \,  \slashed{n} \,\,  u_{j^{\prime}}^{\downarrow}(\tau_2 \, n)[\tau_2 \, n,  \, \tau_0 \, n]_{j^{\prime} j}  \right ] \,
\slashed{n} \,  d_{k^{\prime}}^{\uparrow}(\tau_3 \, n)[\tau_3 \, n,  \, \tau_0 \, n]_{k^{\prime} k}
\right | N^{\uparrow} (P) \right  \rangle
\nonumber \\
&& = - {1 \over 2} \, (n \cdot P)  \, \slashed{n} \,N^{\uparrow}(P) \,
\int [{\cal D} x] \, {\rm exp} \left [ - \, i \, n \cdot P \,  \sum_{i=1}^{3} \, x_i \, \tau_i \right ] \, \Phi_N(x_i, \mu)  \,,
\label{definiton:  the nucleon DA}
\end{eqnarray}
\end{widetext}
by employing the chiral quark fields $q^{\uparrow (\downarrow)} = {1 \over 2} \, (1 \pm \gamma_5)  \, q$
and by introducing   the finite-length  Wilson line  $[\tau_i \, n,  \, \tau_0 \, n]$  to maintain  gauge invariance.
In comparison with the UV renormalization for the two-particle non-local operators \cite{Sarmadi:1982yg,Dittes:1983dy,Katz:1984gf,Mikhailov:1984ii,Belitsky:1999gu,Braun:2017cih,Ji:2023eni,Lange:2003ff,Braun:2019wyx,Liu:2020ydl},
we are now required to enlarge the three-particle collinear operator basis to include  the evanescent operators
\begin{eqnarray}
{\cal O}_{1}  &=&     [  u^{\uparrow}  C \slashed{n} u^{\downarrow} ] \,\,  \slashed{n}   d^{\uparrow},
\nonumber \\
{\cal O}_{2}  &=&     [  u^{\uparrow}  C  \gamma_{\perp \alpha}  \gamma_{\perp \beta} \, \slashed{n} u^{\downarrow} ] \,\,
\slashed{n}  \gamma_{\perp}^{\beta}  \gamma_{\perp}^{\alpha}  d^{\uparrow},
\nonumber \\
{\cal O}_{3}  &=&     [  u^{\uparrow}  C  \gamma_{\perp \alpha}  \gamma_{\perp \beta} \gamma_{\perp \rho}  \gamma_{\perp \tau}   \, \slashed{n} u^{\downarrow} ] \,\,
\slashed{n}  \gamma_{\perp}^{\tau} \gamma_{\perp}^{\rho}    \gamma_{\perp}^{\beta}  \gamma_{\perp}^{\alpha}  d^{\uparrow},
\hspace{0.5 cm}
\label{collinear operator basis}
\end{eqnarray}
for the sake of subtracting all the divergences of the bare matrix element for the physical operator.
It becomes evident that the two evanescent operators ${\cal O}_{2}$ and ${\cal O}_{3}$ vanish at $D=4$
and ${\cal O}_{1}$ is the unique physical operator.
In order to reduce our notation to the essentials, we strip off all the colour indices,
Wilson lines and position arguments from the  light-ray baryonic  operators
${\cal O}_{i} \in \{{\cal O}_{1},  {\cal O}_{2},  {\cal O}_{3} \}$.

The renormalized momentum-space  physical operator  can  be readily expressed
in terms of the three bare operators  at two-loop order
\begin{eqnarray}
{\cal \widetilde{O}}_{1}^{\rm ren}(x_i, \mu) = \sum_{k=1, 2, 3} \int [{\cal D} x^{\prime}] \,
\mathbb{Z}_{1 k}(x_i, x_i^{\prime}, \mu)  \,  \widetilde{O}^{\rm bare}_{k}(x_i^{\prime}) \,,
\hspace{0.5 cm}
\label{definitions: renormalized operators}
\end{eqnarray}
where ${\cal \widetilde{O}}_{i}$  represents  the three-dimensional Fourier transform
of the position-space operator ${\cal O}_{i}$ displayed in  (\ref{collinear operator basis}).
The integration measure is explicitly defined as
$\int [{\cal D} z] = \int_0^1 d z_1 \,  d z_2 \, d z_3 \, \delta(z_1 + z_2 + z_3 -1)$.
Renormalizing the matrix elements of the evanescent operators to zero \cite{Dugan:1990df,Herrlich:1994kh,Buras:1989xd}
enables   us to  write down  the following evolution  equation
\begin{eqnarray}
{d \over d \ln \mu} {\cal \widetilde{O}}_{1}^{\rm ren}(x_i, \mu)
+ \int [{\cal D} x^{\prime}] \,\, \mathbb{H}(x_i,  x_i^{\prime}, \mu) \,
{\cal \widetilde{O}}_{1}^{\rm ren}(x_i^{\prime}, \mu)  =  0,
\hspace{0.5 cm}
\label{RG equation of the nucleon DA}
\end{eqnarray}
where the anomalous dimension  is determined by
\begin{eqnarray}
&& \mathbb{H}(x_i,  x_i^{\prime}, \mu)
\nonumber \\
&& = \sum_{k=1, 2, 3} \int [{\cal D} x^{\prime \prime}]  \, \mathbb{Z}_{1 k}(x_i, x_i^{\prime \prime}, \mu) \,
{d \mathbb{Z}_{k 1}^{-1}(x_i^{\prime \prime}, x_i^{\prime}, \mu)  \over d \ln \mu} \,.
\end{eqnarray}
Performing the double expansion of the renormalization constants
in powers of the strong coupling constant $\alpha_s$ and of their $\epsilon$  poles
in the standard $\overline{{\rm MS}}$ scheme
\begin{eqnarray}
&& \mathbb{Z}_{i j}(x_i, x_i^{\prime}, \mu) =  \sum_{m=0}^{\infty} \,
\left ({\alpha_s(\mu) \over 4 \pi} \right )^{m} \, \mathbb{Z}_{i j}^{(m)}(x_i, x_i^{\prime})
\nonumber  \\
&& \mathbb{Z}_{i j}^{(m)}(x_i, x_i^{\prime}) = \sum_{n=0}^{m} \,
 \left ({1 \over \epsilon} \right )^{n} \,
\mathbb{Z}_{i j}^{(m, n)}(x_i, x_i^{\prime}) \,,
\end{eqnarray}
we are then led to the ``master formula" for the RG evolution kernel  at NLO \cite{Chetyrkin:1997fm}
\begin{eqnarray}
\mathbb{H} &=&   \left ({\alpha_s \over 4 \pi} \right ) \, \mathbb{H}^{(0)}
+  \left ({\alpha_s \over 4 \pi} \right )^{2} \, \mathbb{H}^{(1)}  +  {\cal O}(\alpha_s^2) \,,
\nonumber  \\
\mathbb{H}^{(0)} &=& 2 \, \mathbb{Z}_{1 1}^{(1, 1)} \,,
\nonumber  \\
\mathbb{H}^{(1)} &=&  4  \, \mathbb{Z}_{1 1}^{(2, 1)} - 2  \, \mathbb{Z}_{1 2}^{(1, 1)} \otimes  \, \mathbb{Z}_{2 1}^{(1, 0)}.
\label{master formula for the RG kernel}
\end{eqnarray}
Unsurprisingly,  the  emerged finite renormalization constant $\mathbb{Z}_{2 1}^{(1, 0)}$ due to the evanescent-to-physical operator mixing
is crucial for correctly determining the two-loop anomalous dimension $\mathbb{H}^{(1)}$ with dimensional regularization,
in spite of the vanishing of the evanescent operator ${\cal \widetilde{O}}_{2}$ in four dimensions.
It remains interesting  to point out  that the NLO evolution  kernel $\mathbb{H}^{(1)}$ actually
does not depend on the particular choice of the  evanescent operator ${\cal \widetilde{O}}_{3}$,
on account of the cancellation of the scheme dependence between the two individual pieces
in (\ref{master formula for the RG kernel}).
The required  renormalization factors $\mathbb{Z}_{i j}^{(m, n)}$  can be extracted by computing
the  three-quark matrix elements of the collinear operators
$\Pi_{i} = \langle 0 | {\cal \widetilde{O}}_{i}(x_1, x_2, x_3) |
u^{\uparrow}(P_1) u^{\downarrow}(P_2) d^{\uparrow}(P_3) \rangle $  in perturbation theory,
where the external parton momenta are taken to be $P_i = x_i^{\prime} \, P$ at the leading-power accuracy.
We will perform the two-loop computation of  these QCD  matrix elements with dimensional regularization to capture only the UV divergences
and with the non-vanishing  mass $m_{\rm IR}$ for all internal quarks/gluons regulating the infrared (IR) singularities.

%
\section{The Two-Loop  Evolution Kernel}
%

\begin{figure}[tp]
\begin{center}
\includegraphics[width=0.90 \columnwidth]{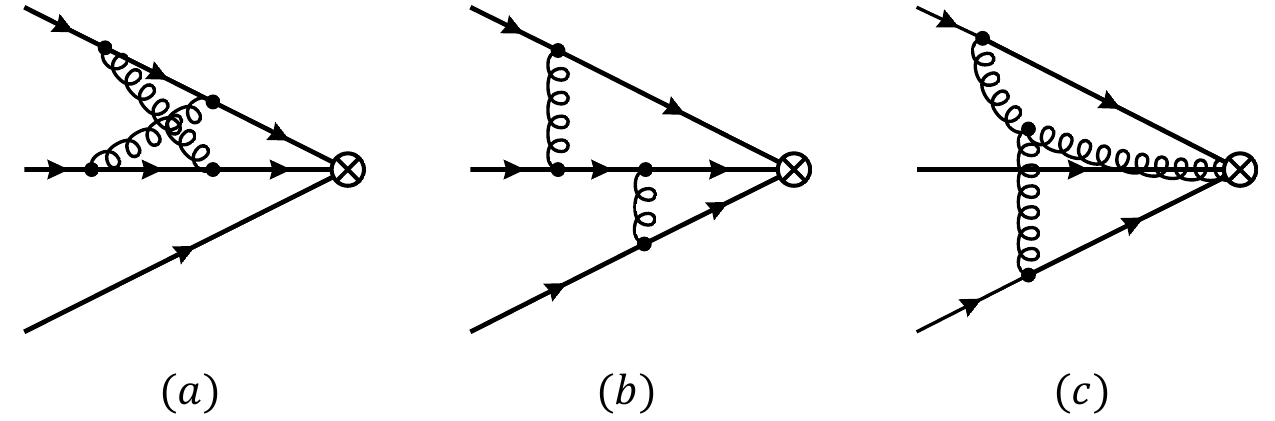}
\caption{Sample two-loop Feynman diagrams for the QCD matrix element $\Pi_{1}$.
The circled cross marks an insertion of the three finite-distance collinear Wilson lines
in (\ref{definiton:  the nucleon DA}).}
\label{fig: sample two-loop Feynman diagrams}
\end{center}
\end{figure}

We are now in a position to describe briefly the  two-loop calculation of the QCD  matrix element
for the physical operator $\Pi_{1}$ at ${\cal O}(\alpha_s^2)$, which allows for the perturbative determination of
the renormalization factor $\mathbb{Z}_{1 1}^{(2, 1)}$.
We first generate the entire set of $70$ Feynman diagrams contributing to $\Pi_{1}$ in a general covariant gauge
by means of  an in-house Mathematica routine.
It can then be observed that  the  peculiar  three-gluon-vertex diagram with the collinear  gluons
emanating from three distinct external quark fields  cannot contribute due to the zero colour factor.
We further note that a subset of the  two-loop  diagrams  that  contain  the  one-loop subgraph
arising  from  the collinear gluon exchange between  two external quarks with the same chirality
can only bring about the UV finite  contributions.
Three sample  Feynman diagrams for $\Pi_{1}$  are explicitly displayed
in Figure \ref{fig: sample two-loop Feynman diagrams}.

Subsequently, we apply the Passarino-Veltman decomposition \cite{Passarino:1978jh} for the vector and tensor integrals
and perform  the Dirac and colour algebraic reduction with in-house  Mathematica routines
based upon the QCD equations of motion and on-shell conditions.
The resulting  two-loop scalar integrals are further reduced to a small set of master integrals
by taking advantage of the  identity
\begin{eqnarray}
&& 2 \, \pi \, i \, \delta \left ( (x_i - x_i^{\prime}) \, n \cdot p - n \cdot \ell \right )
 \\
&& = \frac{1}{(x_i - x_i^{\prime}) \, n \cdot p - n \cdot \ell   -  i 0}
- \frac{1}{(x_i - x_i^{\prime}) \, n \cdot p - n \cdot \ell   +  i 0}
\nonumber
\end {eqnarray}
and implementing  the integration-by-parts relations \cite{Tkachov:1981wb,Chetyrkin:1981qh}
and the Laporta algorithm \cite{Laporta:1996mq,Laporta:2000dsw}.
To this end,  we use version 6 of {\tt FIRE} \cite{Smirnov:2019qkx} in combination with an in-house routine.
Exploiting the very fact that the three external quarks move in nearly the same direction $\bar n$,
we then arrive at a set of  $20$  two-loop  master integrals in our computation.
It turns out that  $13$  master integrals already  appeared in the  NLO  calculation
of the  renormalization kernel for the twist-two pion LCDA \cite{Mikhailov:1984ii,Onishchenko:2003hs}.
The UV divergent contributions of all  master integrals  can be analytically  evaluated  with the residue theorem
and can be further  expressed in terms of the usual polylogarithms up to weight $2$.
Along the same vein, we can proceed to  determine the two remaining renormalization constants
$\mathbb{Z}_{1 2}^{(1, 1)}$ and $\mathbb{Z}_{2 1}^{(1, 0)}$ by computing the QCD matrix elements $\Pi_{1, 2}$
at the one-loop level.
The obtained  expression of the substraction term $\mathbb{Z}_{1 2}^{(1, 1)} \otimes  \, \mathbb{Z}_{2 1}^{(1, 0)}$
characterizing  the evanescent-to-physical operator mixing   is explicitly  presented in the Supplemental Material.

Having at our disposal the desired  results for the necessary  one- and two-loop  renormalization factors,
we can now  derive the NLO renormalization kernel of the leading-twist nucleon distribution amplitude $\Phi_N$
with the aid of the master formula (\ref{master formula for the RG kernel})
\begin{widetext}
\begin{eqnarray}
\mathbb{H}^{(1)} &=&  \bigg \{  \left ( {C_F  \over N_c -1} \right ) \, C_A \,
\left  [ \mathbb{V}^{(1), \, C_F \, C_A}_{\rm LC}(\chi_i) \,
\delta(x_1 - x_1^{\prime}) \, \delta(x_2 - x_2^{\prime})
+  \,  {\color{blue} \mathbb{V}^{(1), \, C_F \, C_A}_{\rm 2P}(x_1, x_2, x_1^{\prime}, x_2^{\prime}, \chi_i) }
\, \delta(x_3 - x_3^{\prime})  \right  ]
\nonumber \\
&& +  \,  \left ( {C_F  \over N_c -1} \right ) \, \beta_0 \,
\left [  \mathbb{V}^{(1), \, C_F \, \beta_0}_{\rm LC}(\chi_i) \,
 \delta(x_1 - x_1^{\prime}) \, \delta(x_2 - x_2^{\prime})
 +  \,  {\color{blue} \mathbb{V}^{(1), \, C_F \, \beta_0}_{\rm 2P}(x_1, x_2, x_1^{\prime}, x_2^{\prime}, \chi_i)}
\, \delta(x_3 - x_3^{\prime})   \right ]
\nonumber \\
&& + \, \left ( {C_F  \over N_c -1} \right )^2 \,
\left [ \mathbb{V}^{(1), \, C_F^2}_{\rm LC}(\chi_i) \,
\delta(x_1 - x_1^{\prime}) \, \delta(x_2 - x_2^{\prime})
+  \,  {\color{blue} \mathbb{V}^{(1), \, C_F^2}_{\rm 2P}(x_1, x_2, x_1^{\prime}, x_2^{\prime}, \chi_i) }
\, \delta(x_3 - x_3^{\prime}) \right .
\nonumber \\
&& \hspace{2.5 cm}    \left . {\color{magenta} +  \, \mathbb{V}^{(1), \, C_F^2}_{\rm 3P}(x_i, x_i^{\prime}, \eta_i, \kappa_i, \chi_i)}  \right  ]  \bigg \}
+ \,  \left \{ x_1 \leftrightarrow x_3, \, \chi_i \leftrightarrow \eta_i   \right \}
+  \left \{ x_2 \leftrightarrow x_3, \, \chi_i  \leftrightarrow \kappa_i \right \}.
\label{master formula of the NLO evolution kernel}
\end {eqnarray}
\end{widetext}
The appearance of three diagonal terms in (\ref{master formula of the NLO evolution kernel})
with the  structure  $\delta(x_1 - x_1^{\prime}) \, \delta(x_2 - x_2^{\prime})$
manifests the fact that the  {\it local} three-particle  operator
$ \epsilon_{i j k} \, [  u_{i}^{\uparrow}(0) \,  C \slashed{n} \, u_{j}^{\downarrow}(0) ] \,\,  \slashed{n}   \, d_{k}^{\uparrow}(0)$
is perturbatively renormalized in QCD.
It is straightforward to  verify that the two-loop  evolution  kernel of the leading-twist pion distribution amplitude
\cite{Sarmadi:1982yg,Dittes:1983dy,Katz:1984gf,Mikhailov:1984ii,Belitsky:1999gu}
can be fully  constructed  from (\ref{master formula of the NLO evolution kernel}) with the following  procedure:
I) discarding the symmetric terms $\left \{ x_1 \leftrightarrow x_3, \, \chi_i \leftrightarrow \eta_i   \right \}$
and $\left \{ x_2 \leftrightarrow x_3, \, \chi_i \leftrightarrow  \kappa_i   \right \}$
due to permutations of the three quark flavours,
II) implementing   the obvious replacement for  colour factors  $C_F / (N_c -1) \to C_F$,
III)  retaining only  the  two-particle   interaction effects $\mathbb{V}^{(1), \, C_F \, C_A}_{\rm 2P}$,
$\mathbb{V}^{(1), \, C_F \, \beta_0}_{\rm 2P}$ and $\mathbb{V}^{(1),  \, C_F^2}_{\rm 2P}$
(but excluding the peculiar term  $\Delta \mathbb{V}^{(1),  \, C_F^2}_{\rm 2P}$ therein).
In  addition, we  can  continue to   reproduce in an analogous fashion the NLO renormalization kernel
of the leading-twist distribution amplitude   for the transversely polarized $\rho$-meson $\Phi_{\perp}$  \cite{Erratum-rho-meson-RGE},
which has been previously determined  with  the conformal symmetry technique \cite{Belitsky:2000yn}
and  with the  diagrammatic approach \cite{Mikhailov:2008my}.
Applying  further the two-loop conversion factor for  the nucleon distribution amplitude $\Phi_{N}$
between our renormalization scheme with the presence of evanescent operators (hereafter ``the EO scheme")
and the KM scheme
\begin{eqnarray}
\Phi_N^{\rm KM}(x_i, \mu) = \int [{\cal D} x_i^{\prime}] \,\, \mathbb{K}_{N}(x_i,  x_i^{\prime}, \mu)
\,\,  \Phi_N^{\rm EO}(x_i^{\prime}, \mu),
\label{conversion factor}
\hspace{0.7 cm}
\end{eqnarray}
we can readily confirm   the  available  two-loop  anomalous dimensions 
for both the normalization coefficient and the first three shape parameters
of the leading-twist nucleon distribution amplitude with the KM  renormalization prescription  \cite{Krankl:2011gch,Bali:2024oxg},
thus providing  non-trival checks  of  the newly obtained  NLO evolution kernel $\mathbb{H}^{(1)}$.
The manifest expressions for the primitive kernels $\mathbb{V}^{(1), \, n}_{\rm LC}$, $\mathbb{V}^{(1), \, n}_{\rm 2P}$
and $\mathbb{V}^{(1), \, n}_{\rm 3P}$ (with $n=C_F \, C_A,  \, C_F  \, \beta_0, \, C_F^2$)
entering the NLO renormalization kernel  (\ref{master formula of the NLO evolution kernel})
and for  the coefficient function $\mathbb{K}_{N}$ at the two-loop accuracy
are  displayed  in the Supplemental Material for completeness.

%
\section{The Analytic Solution}
%

We are now prepared to determine  the  scale dependence of the leading-twist nucleon distribution amplitude
by  solving the  integro-differential evolution  equation  (\ref{RG equation of the nucleon DA})
with the inclusion of the two-loop RG  kernel $\mathbb{H}^{(1)}$.
It turns out to be advantageous to apply  the conformal partial wave expansion
of the nucleon distribution amplitude $\Phi_N$ in terms of  orthogonal polynomials ${\cal P}_{M m}$
defined as eigenfunctions of the LO  evolution  kernel $\mathbb{H}^{(0)}$
\begin{eqnarray}
&& \Phi_N(x_i, \mu)
\nonumber \\
&& =  x_1 \,  x_2 \,  x_3 \, \sum_{M=0}^{\infty} \, \sum_{m=0}^{M} \,
{\cal N}_{M m} \, \Psi_{M m}(\mu) \, {\cal P}_{M m}(x_i),
\label{conformal expansion of the nucleon DA}
\hspace{0.75 cm}
\end{eqnarray}
where the  normalization coefficients ${\cal N}_{M m} $ are  determined  from
$ \int [{\cal D} x] \,  x_1 \, x_2 \, x_3 \, [ {\cal P}_{M m}(x_i) ]^2 = {\cal N}_{M m}^{-1}$.
We can then  readily translate the RG equation  (\ref{RG equation of the nucleon DA})  of the nucleon distribution amplitude
into the following evolution equation of the local moments $\Psi_{Q q}$
\begin{eqnarray}
\sum_{Q=0}^{M}   \sum_{q=0}^{Q}  \left [ {d \over d \ln \mu}  \delta_{M Q}  \delta_{m q}
+  \bbGamma_{M m, \,  Q q}   (\mu) \right ]  \Psi_{Q q}  (\mu) = 0,
\hspace{0.75 cm}
\label{RG equation for the local moments}
\end{eqnarray}
where the anomalous  dimension matrix  $\bbGamma$ is  given by
\begin{eqnarray}
\bbGamma_{M m,  \, Q q}  (\mu)  
&=& {\cal N}_{Q q} \, \int [{\cal D} x] \, \int [{\cal D} x^{\prime}] \,
x_1^{\prime} \,  x_2^{\prime} \,  x_3^{\prime} \,
\nonumber \\
&&  \times \, \left [  {\cal P}_{M m}(x_i) \,\,  \mathbb{H}(x_i,  x_i^{\prime}, \mu) \,\,  
{\cal P}_{Q q}(x_i^{\prime})  \right ].
\hspace{0.8 cm}
\end{eqnarray}
The general solution to this  ordinary  differential equation  can be cast the form of
\begin{eqnarray}
\Psi_{M m}  (\mu) &=& \sum_{Q=0}^{M}   \sum_{q=0}^{Q}   \,\,
\left [ \mathbb{U}(\mu, \, \mu_0) \right ]_{M m, \,  Q q} \, \Psi_{Q q}  (\mu_0)\,,
\nonumber \\
 \mathbb{U}(\mu, \, \mu_0)  &=&  \mathbb {T}_{\mu} \,
 {\rm exp} \left [ - \int_{\mu_0}^{\mu} \, { d \nu  \over \nu }\, \bbGamma(\nu) \right ] \,,
\end{eqnarray}
where the $\mu$-ordering operator $\mathbb {T}_{\mu}$  is  defined as \cite{Buras:1979yt,Buras:1998raa}
\begin{eqnarray}
&& \mathbb {T}_{\mu} \,  f(\mu_1) \, ... \, f(\mu_n)
\nonumber \\
&& = \sum_{\rm perm} \, \theta(\mu_{i_1} - \mu_{i_2}) \,  ... \, \theta(\mu_{i_{n-1}} - \mu_{i_{n}})
\, f(\mu_{i_1}) \, ...  \, f(\mu_{i_{n}}).
\hspace{0.7 cm}
\end{eqnarray}
Keeping the first two terms in the perturbative expansions of $\bbGamma$ and
the QCD $\beta$-function allows us to derive further  the evolution matrix in the NLL approximation
\begin{eqnarray}
\mathbb{U}^{\rm NLL}(\mu, \, \mu_0) &=& \left ( \mathbb{I}  +  {\alpha_s(\mu) \over 4 \, \pi}   \mathbb{J}^{(1)} \right )
\left [ \left  (  \frac{\alpha_s(\mu)}   {\alpha_s(\mu_0)} \right) ^{\bbGamma^{(0)} \over 2 \, \beta_0 } \right ]
\nonumber \\
&& \, \times \,  \left ( \mathbb{I}  -  {\alpha_s(\mu_0) \over 4 \, \pi} \, \mathbb{J}^{(1)} \right ),
\label{NLL evolution function}
\end{eqnarray}
where  the matrix $\mathbb{J}^{(1)}$ has the entries
\begin{eqnarray}
&& \mathbb{J}_{M m, \, Q q}^{(1)}
\nonumber \\
&& = { \beta_1  \over \beta_0} \,  \frac{\bbGamma^{(0)}_{M m, \, Q q}} {2 \, \beta_0}
 -   \frac{\bbGamma^{(1)}_{M m, \, Q q}} {2 \, \beta_0 - \bbGamma^{(0)}_{M m, \, M m} + \bbGamma^{(0)}_{Q q, \, Q q}} \,.
\end{eqnarray}
We  present  explicitly the  analytic expressions for the anomalous dimension matrices $\bbGamma^{(0), \, (1)}$
and  for the  orthogonal  polynomials ${\cal P}_{M m}$ in the Supplemental Material
by truncating the conformal expansion (\ref{conformal expansion of the nucleon DA}) at $M=3$,
which is evidently sufficient for practical purposes \cite{Anikin:2013aka,Huang:2024ugd}.

%
\section{Numerical Implications}
%

We will dedicate this section to investigating   phenomenological implications  of the newly  derived
two-loop RG  evolution of the leading-twist nucleon distribution amplitude.
It is then instructive to employ  three nonperturbative models for the initial condition
$\Phi_{N}(x_i, \, \mu_0)$ at a reference scale $\mu_0 = 1.0 \, {\rm GeV}$,
labelled as {\tt COZ} \cite{Chernyak:1987nt},  {\tt LAT25} \cite{Bali:2024oxg},
and {\tt ABO1} \cite{Anikin:2013aka}, in the subsequent numerical analysis.
The distinctive feature of the classic {\tt COZ} model
(motivated by  QCD sum  rule  estimates  of the  ten lowest  moments)
consists in an  enormously large fraction of the  proton  momentum
carried by the first $u$-quark with the same helicity:
approximately $60 \%$ in the collinear limit.
The  second parameter set ${\tt LAT25}$ is determined from  the ab initio lattice QCD  calculation
with $N_f= 2 + 1$ flavours of dynamical Wilson fermions
and  further improved by incorporating  the  two-loop conversion factors
for local three-quark operators between the momentum-subtraction scheme and the ${\rm \overline{MS}}$  scheme \cite{Krankl:2011gch}.
By  contrast, the construction of the sample  model  {\tt ABO1} is achieved
by matching the state-of-the-art  light-cone sum rule predictions of the nucleon electromagnetic form factors \cite{Anikin:2013aka}
onto the available experimental measurements \cite{Arrington:2007ux,CLAS:2008idi,JeffersonLabHallA:2001qqe,Punjabi:2005wq,JeffersonLaboratoryE93-038:2005ryd,Riordan:2010id}.
It is perhaps worth mentioning that another  alternative and model-independent technique
of extracting the whole  profile of the twist-three light-baryon distribution amplitude
has been recently proposed \cite{Deng:2023csv,Han:2023xbl,Han:2023hgy,Han:2024ucv,LatticeParton:2024vck,LPC:2025jvd}
by performing  the numerical simulation of an appropriate time-independent
quasi-distribution function in the framework of large momentum effective theory \cite{Ji:2013dva,Ji:2014gla}.

\begin{figure}[htp]
\begin{center}
\includegraphics[width=0.90 \columnwidth]{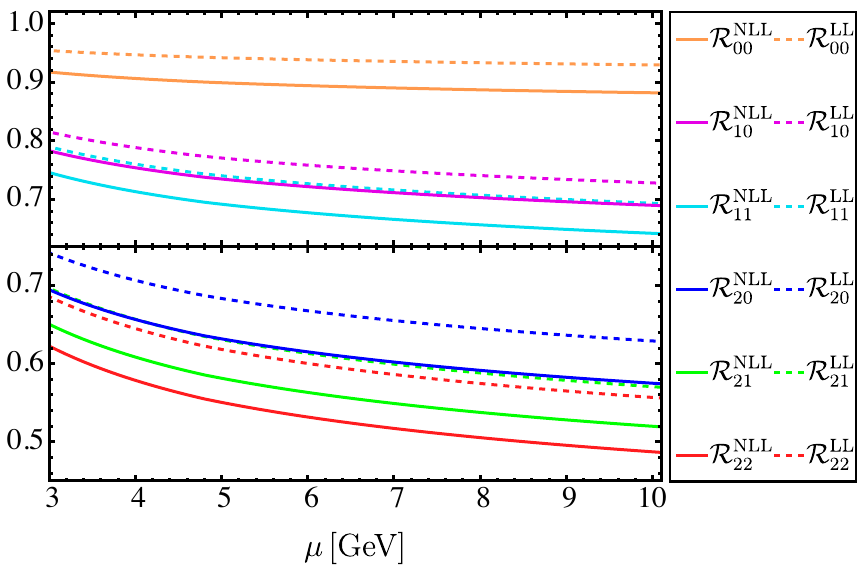}
\caption{Theory predictions for the RG evolution of the normalized moments ${\cal R}_{M m}$
(with $M=0, \, 1,  \, 2$) of the leading-twist nucleon distribution amplitude at the LL order (dotted curves)
and at the  NLL order (solid curves) in QCD,  obtained from the sample  {\tt COZ}   model
of $\Phi_{N}(x_i, \, \mu_0)$.}
\label{fig: RG Evolution of the moments in the COZ model}
\end{center}
\end{figure}

In order to develop a transparent understanding of the numerical feature
for  the two-loop RG evolution of the nucleon distribution amplitude,
we display explicitly in Figure \ref{fig: RG Evolution of the moments in the COZ model}
the  yielding  predictions for  the normalized  shape parameters
${\cal R}_{M m}(\mu, \, \mu_0) = \Psi_{M m}(\mu) : \Psi_{M m}(\mu_0)$ with  $M  \leq  2$
in the range of $\mu \in [3.0, \, 10.0] \, {\rm GeV}$
at the leading-logarithmic (LL) order and at the NLL order,
by adopting the {\tt COZ} model of $\Phi_{N}(x_i, \, \mu_0)$  for the purpose of illustration.
It is evident from such comparative explorations  that  the very inclusion of
the newly determined NLL corrections to  the non-perturbative coefficients  $\Psi_{M m}$
can bring about  noticeable impacts on the  corresponding LL predictions at intermediate renormalization scales:
numerically at the level of ${\cal O} (20 \, \%)$.
Remarkably,  the two-loop QCD evolution for the nucleon distribution amplitude
can give rise to far more pronounced effects  than that for the leading-twist  distribution amplitude
of  the $\pi$-meson \cite{Sarmadi:1982yg,Dittes:1983dy,Katz:1984gf,Mikhailov:1984ii,Belitsky:1999gu,Strohmaier:2018tjo}
and of the $B$-meson \cite{Braun:2019wyx,Liu:2020ydl}, thus justifying the  prominent significance of
carrying out the full  NLO  computation  of the RG evolution kernel for $\Phi_{N}$.
We further verify  that this  intriguing  pattern of  the  NLL  evolution of the nucleon distribution amplitude
remains unchanged for the two additional  model functions {\tt LAT25}  and {\tt ABO1}.
It is also  interesting to note that  the considered  moments with higher conformal spins 
receive  more  substantial corrections numerically from the NLL  perturbative  evolution.

\begin{figure}[htp]
\begin{center}
\includegraphics[width=0.90 \columnwidth]{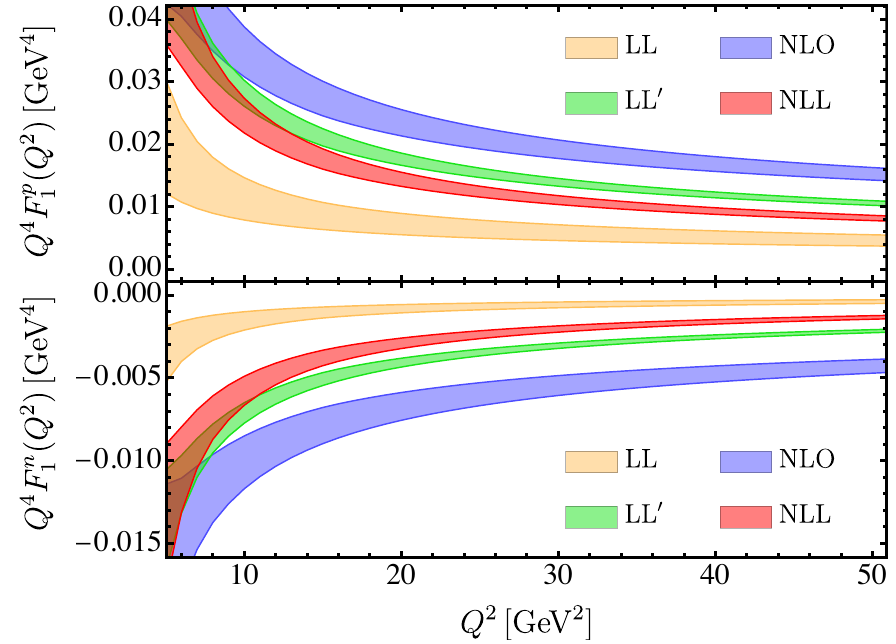}
\caption{Theory predictions for the  leading-power hard-gluon-exchange contributions to
the Dirac nucleon form factors of both the proton (upper panel) and the neutron (lower panel)
at the LL, NLO, ${\rm LL}^{\prime}$ and NLL accuracy
by taking the parameter set ${\tt LAT25}$ for the nucleon distribution amplitude.
The perturbative uncertainties from varying the renormalization and factorization scales
in the preferred intervals,  $\nu^2=\mu^2 = \langle x \rangle \, Q^2$
with $ 1/6 \leq \langle x \rangle \leq 1/2$ \cite{Huang:2024ugd},  are indicated by the colour bands. }
\label{fig: Numerics of the nucleon form factors with the LAT25 model}
\end{center}
\end{figure}

We finally address the  genuine  impact of the two-loop QCD evolution of $\Phi_{N}$
on the hard-scattering contributions to the Dirac nucleon form factors
computed from the hard-collinear factorization  formalism.
To  this end, we present in Figure \ref{fig: Numerics of the nucleon form factors with the LAT25 model}
the obtained theory  predictions for these fundamental hadronic quantities with the particular  model ${\tt LAT25}$
in the LL, NLO, ${\rm LL}^{\prime}$ and NLL  approximations,
where the  ${\rm LL}^{\prime}$ accuracy is routinely  defined by
including the fixed NLO correction in the LL resummation improved contribution
(see for instance \cite{Abbate:2010xh,Berger:2010xi,Almeida:2014uva}).
The distinctive snapshot of the well-separated uncertainty bands  in the kinematic domain
$10.0 \, {\rm GeV^2} \leq Q^2 \leq 50.0 \, {\rm GeV^2}$  from  perturbative QCD calculations
at the four different orders  unequivocally  elucidates the profound significance
of taking into account the two-loop RG evolution of the leading-twist nucleon distribution amplitude.
We further observe that the  predicted NLL  corrections to the Dirac nucleon form factors
appear to  become numerically more  important for  higher momentum transfers.
In particular, the achieved  theory  predictions  for the Dirac neutron form factor $F_1^{n}(Q^2)$
are  considerably  more  affected by the NLL QCD  resummation effects
when compared with  the  determined  results for  the counterpart proton  form factor $F_1^{p}(Q^2)$,
confirming  an  earlier conjecture on the extraordinarily sizeable two-loop  radiative corrections to
the neutron electromagnetic form factors \cite{Anikin:2013aka}.

\section{Conclusions}

In  summary,  we have endeavored to compute  for the first time  the NLO  evolution kernel
of the leading-twist nucleon distribution amplitude $\Phi_{N}$ in QCD
by virtue of the modern effective field theory formalism.
Taking advantage of the conformal partial wave expansion for  $\Phi_{N}$ then allowed us to determine in an analytic fashion
the desired scale dependence of the normalization constant and  shape parameters at NLL.
Equipped with the  thus  derived  two-loop QCD  evolution of  conformal moments, we  presented further
the complete NLL predictions of the leading-twist contributions to the Dirac nucleon form factors
by adopting three sample models  for the nucleon distribution amplitude.
It has been demonstrated that  the newly obtained NLL corrections to  $\Phi_{N}$ can result in  significant impacts
on the predicted nucleon form factors over a wide range of momentum transfers.
Extending our RG analysis to the leading-twist distribution amplitudes  of the full  baryon octet and decuplet
will  be  highly beneficial  for  exploring  the mysterious  partonic landscape of these composite hadron systems
and for achieving the precision QCD  description of  a large variety of hard exclusive reactions
(such as the electroproduction of the $\Delta$-resonance,  the weak radiative hyperon decay,
and the electroweak penguin $\Lambda_b \to \Lambda \ell^{+} \ell^{-}$ decay).

%
\begin{acknowledgments}
\section*{Acknowledgements}

We are grateful to  Alexander Manashov for illuminating discussions and
for  valuable  comments/suggestions  on the manuscript.
Y.K.H., B.X.S. and  Y.M.W. acknowledges support from the  National Natural Science Foundation of China
with Grants No. 12475097 and No. 12535006, and from  the Natural Science Foundation of Tianjin
with Grant No. 25JCZDJC01190.
The  research of Y.J. is supported in part by the National Natural Science Foundation
of China with Grant No.12535006, and by the University Development Fund
of the Chinese University of Hong Kong, Shenzhen under the grant No. UDF0100386.

\end{acknowledgments}

%
\appendix

\begin{widetext}

\section{SUPPLEMENTAL MATERIAL}

\subsection{Analytic Expressions for  the  Evolution Kernels}

We collect here the analytic expressions for the  necessary  ingredients  appearing in
the  renormalization kernel $\mathbb{H}$ for the leading-twist nucleon distribution amplitude  $\Phi_{N}$.
The well-known LO evolution kernel $\mathbb{H}^{(0)}$ in QCD \cite{Lepage:1979za,Lepage:1980fj} can be cast in the form of
\begin{eqnarray}
\mathbb{H}^{(0)} &=& \left ( {4 \, C_F \over N_c - 1} \right )
\Bigg \{ \left [  {x_1 \over x_1^{\prime}}
\left (  {1 \over x_1  - x_1^{\prime}}  - {1 \over x_1^{\prime} + x_2^{\prime}} \right )
\theta(x_1^{\prime} - x_1)
+  {x_2 \over x_2^{\prime}}
\left (  {1 \over x_2  - x_2^{\prime}}  - {1 \over x_1^{\prime} + x_2^{\prime}} \right )
\theta(x_2^{\prime} - x_2)  \right ]_{+}
\delta \left ( \sum_{k=1}^{2}   [x_k - x_k^{\prime}] \right )
\nonumber \\
&& \hspace{2.0 cm}  + \,  \left [ x_1 \leftrightarrow x_3, \, x_1^{\prime} \leftrightarrow x_3^{\prime} \right ]
+ \,  \left [  x_2 \leftrightarrow x_3, \, x_2^{\prime} \leftrightarrow x_3^{\prime} \right ]  \Bigg \}
\nonumber \\
&& { \color{magenta}  + \, \left ( {4 \, C_F \over N_c - 1} \right ) \,  \left [ {x_1 \over x_1^{\prime}}  \,
\frac{\theta(x_1^{\prime} - x_1) } {x_1^{\prime} + x_3^{\prime}}
+  {x_3 \over x_3^{\prime}}  \,
\frac{\theta(x_3^{\prime} - x_3) } {x_1^{\prime} + x_3^{\prime}} \right ]_{+}    \,
\delta(x_2 - x_2^{\prime}) } \,
+ \, \left ( {2 \, C_F \over N_c - 1} \right )  \, \delta(x_1 - x_1^{\prime})  \, \delta(x_2 - x_2^{\prime})   \,,
\label{LO evolution kernel}
\end{eqnarray}
where we have introduced  the  modified  definition of the  ``$+$" distribution in the two variables $x_{1, 2}$
\begin{eqnarray}
\left [f(x_1, \, x_2, \,  x_1^{\prime}, \, x_2^{\prime}) \right ]_{+}  \equiv
f(x_1, \, x_2, \,  x_1^{\prime}, \, x_2^{\prime})
- \delta(x_1 - x_1^{\prime}) \, \int_0^1 d y_1 \, \int_0^1 d y_2 \,\,
\delta \left ( \sum_{k=1}^{2}   \, [x_k^{\prime} - y_k] \right ) \,\,
f(y_1, \, y_2, \,  x_1^{\prime}, \, x_2^{\prime})  \,.
\end{eqnarray}
We now turn to present the manifest  expressions for the primitive kernels
$\mathbb{V}^{(1), \, n}_{\rm LC}$, $\mathbb{V}^{(1), \, n}_{\rm 2P}$
and $\mathbb{V}^{(1), \, n}_{\rm 3P}$  entering the newly achieved result (\ref{master formula of the NLO evolution kernel})
of the  NLO  renormalization  kernel $\mathbb{H}^{(1)}$
\begin{eqnarray}
\mathbb{V}^{(1), \, C_F \, C_A}_{\rm LC}&=&  {38 \over 3} - {11 \over 3} \, \chi_1 + {1 \over 8} \, \chi_2 \,,
\qquad \hspace{2.0 cm}
\mathbb{V}^{(1), \, C_F \, \beta_0}_{\rm LC} = {13 \over 3} - {1 \over 12} \, \chi_1 - {1 \over 2} \, \chi_4   \,,
\nonumber \\
\mathbb{V}^{(1), \, C_F^2}_{\rm LC} &=&  -19 + 6  \, \chi_1 -  {5 \over 16} \, \chi_2  \,
{ \color{magenta}  + \, \Delta \mathbb{V}^{(1), \, C_F^2}_{\rm LC} } \,,
\qquad
\Delta \mathbb{V}^{(1), \, C_F^2}_{\rm LC} = -11 + {5 \over 2}  \, \chi_1 -  {1 \over 8} \, \chi_3  \,,
 \\
\nonumber \\
\mathbb{V}^{(1), \, C_F \, C_A}_{\rm 2P} &=&
\Bigg [ {16 \over 3}   { x_1 \over x_1^{\prime}} \, \left ( {1 \over x_1 - x_1^{\prime}}
- {5 \over 8}  {1 \over x_1^{\prime} +  x_2^{\prime} }  \chi_1 \right )  \theta(x_1^{\prime} - x_1)
+ \left (2 - {1 \over 8}  \chi_2 \right )
\left (  {x_1  +  x_2^{\prime}  \over x_1^{\prime}  \, x_2^{\prime} }   \theta(x_1^{\prime} - x_1)
- {  x_2^{\prime} - x_1 \over x_1^{\prime}  \, x_2^{\prime} }  \theta(x_2^{\prime} - x_1) \right )
\nonumber \\
&& - \, \left ( \chi_1 - {1 \over 4} \, \chi_2 \right ) \,
{x_1 \over x_1^{\prime} \, ( x_1^{\prime}  + x_2^{\prime} ) } \,
\left ( \ln{x_1 \over x_2^{\prime} } \, \theta(x_2^{\prime} - x_1)
+ {x_1^{\prime} \over x_2^{\prime}}  \, \ln {x_1 \over x_1^{\prime} } \, \theta(x_1^{\prime} - x_1)
- { x_1^{\prime} + x_2^{\prime} \over x_2^{\prime} } \, \ln { x_1 \over x_1^{\prime} + x_2^{\prime}} \right )
\nonumber \\
&&  - \, { x_1 \over x_1^{\prime}}  \,
\left ( {4 \over x_1 - x_1^{\prime}} - {1 \over x_1^{\prime} +  x_2^{\prime} } \, \chi_1 \right ) \,
{ \color{blue} \Bigg ( }
\left ( 2 \, {\rm Li}_2 \left ( { x_1 \over x_1^{\prime} + x_2^{\prime}} \right )
+ \ln^2 { x_1 \over x_1^{\prime} + x_2^{\prime}}  \right )  \,
\left ( \theta(x_1^{\prime} - x_1) - \theta(x_2^{\prime} - x_1)   \right )
\nonumber \\
&& \hspace{0.3 cm} -  \, \left ( 2 \, {\rm Li}_2 \left ( { x_2^{\prime} \over x_1^{\prime} + x_2^{\prime}} \right ) 
+ \ln^2 { x_2^{\prime} \over x_1^{\prime} + x_2^{\prime}}  \right ) \,
\left ( \theta(x_1 - x_1^{\prime}) - \theta(x_2^{\prime} - x_1)   \right )
+ 2 \,  {\rm Li}_2 \left ( 1 - { x_1^{\prime} \over x_1} \right )  \, \theta(x_1 - x_2^{\prime})
\nonumber \\
&& \hspace{0.3 cm} + \, 2 \,  {\rm Li}_2 \left ( 1 - { x_2 \over x_2^{\prime}} \right )  \, \theta(x_2^{\prime} - x_1)
+ \ln^2 {x_1 \over x_2^{\prime}} \,  \theta(x_1 - x_1^{\prime})  { \color{blue} \Bigg ) } \Bigg ]_{+}
+  \left [ x_1 \leftrightarrow x_2,  \, x_1^{\prime}  \leftrightarrow  x_2^{\prime}  \right ] \,,
\\
\nonumber \\
\mathbb{V}^{(1), \, C_F \, \beta_0}_{\rm 2P} &=&  \Bigg [ { x_1 \over x_1^{\prime} }  \,
 \left (  { 1 \over  x_1^{\prime}  + x_2^{\prime}  }  \,  (\chi_1 - \chi_4 )
+   \left (  {4 \over x_1 - x_1^{\prime}} -  {1 \over x_1^{\prime}  + x_2^{\prime} } \, \chi_1 \right )  \,
\left ( \ln { x_1 \over x_1^{\prime}} + {5 \over 3} \right ) \right )
\, \theta(x_1^{\prime} - x_1)  \, \Bigg ]_{+}
\nonumber \\
&& +  \left [ x_1 \leftrightarrow x_2,  \, x_1^{\prime}  \leftrightarrow  x_2^{\prime}  \,\right ]  \,,
 \\
\nonumber \\
\mathbb{V}^{(1), \, C_F^2}_{\rm 2P} &+& 2 \, \mathbb{V}^{(1), \, C_F \, C_A}_{\rm 2P}
=  \Bigg [  {3 \, x_1  \over x_1^{\prime} \, (x_1^{\prime} + x_2^{\prime})}  \,
\left (  1 + { x_1^{\prime} \over x_2^{\prime} }  \, \ln { x_1^{\prime} \over x_1^{\prime} + x_2^{\prime}}
+ { x_2 \over x_1}  \, \ln { x_2 \over x_1^{\prime} + x_2^{\prime}} \right )  \,
\left ( \chi_1 - {1 \over 6} \, \chi_2 \right ) \, \theta(x_1^{\prime} - x_1)
\nonumber \\
&& + \,  { x_1 \over x_1^{\prime} }  \,
\left ( { 8 \over x_1 -  x_1^{\prime} }  - {2 \over x_1^{\prime} + x_2^{\prime}} \, \chi_1 \right ) \,
{ \color{blue} \Bigg (  } \ln{ x_1 \over x_1^{\prime}} \, \left ( \ln { x_1 \over x_1^{\prime} - x_1}
- {3 \over 2} \right ) \, \theta( x_1^{\prime} - x_1)
- \ln{ x_2 \over x_2^{\prime} }  \, \ln{ x_2^{\prime} \over x_1 -  x_1^{\prime}} \, \theta(x_1  - x_1^{\prime})
\nonumber \\
&& \hspace{0.3 cm}    - \ln { x_1 \over  x_1^{\prime}  +  x_2^{\prime} }  \,
\ln { x_2 \over  x_1^{\prime}  +  x_2^{\prime} }  { \color{blue} \Bigg ) }
+ \, { 2 \, x_1 \over x_2^{\prime} \, (x_1^{\prime}  +  x_2^{\prime} ) }  \,
\left (\chi_1 - {1 \over 8} \, \chi_2 \right )  \,
\left ( \ln^2 { x_1 \over x_1^{\prime} + x_2^{\prime}}
- \ln^2 { x_1  \over x_1^{\prime}}  \,\, \theta(x_1^{\prime} - x_1) \right )
\nonumber \\
&& + \, { x_1 \over x_1^{\prime} } \, \left ( {4 - \pi^2 \over 3}  \, { 8 \over x_1 - x_1^{\prime} }
-   { 29 - 2 \, \pi^2 \over 3 } \, { 1 \over x_1^{\prime} + x_2^{\prime}} \, \chi_1 \right ) \, \theta(x_1^{\prime} - x_1) \,
{ \color{magenta}  + \, \Delta  \mathbb{V}^{(1), \, C_F^2}_{\rm 2P} \, \theta(x_1^{\prime} - x_1) }  \Bigg ]_{+}
\nonumber \\
&& +  \,  \left [ x_1 \leftrightarrow x_2,  \, x_1^{\prime}  \leftrightarrow  x_2^{\prime}  \,\right ]  \,,
\\
\nonumber \\
\Delta  \mathbb{V}^{(1), \, C_F^2}_{\rm 2P} &=&
-  \left [ 2 \, {\rm Li}_2 \left ({ x_1^{\prime} \over x_1^{\prime} + x_2^{\prime} }  \right )
+ 2 \, {\rm Li}_2 \left (   { x_1  \over x_1^{\prime} } \right )
- 2 \, {\rm Li}_2 \left ({ x_1 \over x_1^{\prime} + x_2^{\prime} }  \right )
+ \ln^2 { x_1^{\prime}  \over x_1^{\prime} + x_2^{\prime} }
+ \ln^2 { x_2  \over x_1^{\prime} + x_2^{\prime} }  \right ]
\nonumber \\
&& \hspace{0.3 cm} \times \, \left [  { x_1 \over x_1^{\prime}}  \, \left (  { 4 \over x_1 - x_1^{\prime}}
- {1 \over x_1^{\prime} + x_2^{\prime}} \, \chi_1 \right )
+ { x_2 \over x_2^{\prime}}  \, \left (  { 4 \over x_2 - x_2^{\prime}}
- {1 \over x_1^{\prime} + x_2^{\prime}} \, \chi_1 \right )  \right ]
\nonumber \\
&& + \, {x_1 \over x_1^{\prime}} \, \left ( {4 \over x_1 - x_1^{\prime}}
-  {1 \over x_1^{\prime} + x_2^{\prime}} \, \chi_1 \right ) \,
\left [ 2 \, {\rm Li}_2 \left ( 1 - { x_2 \over x_2^{\prime}} \right )
+ \ln^2 { x_1 \over x_2 }  - 2 \, \ln{ x_1 \over  x_1^{\prime}} \, \ln{ x_1 \over  x_2^{\prime}}
- 3 \, \ln{ x_1 \over  x_1^{\prime}}  \right ]
\nonumber \\
&& - \, { x_2 \over x_2^{\prime}}  \, \left (  { 4 \over x_2 - x_2^{\prime}}
- {1 \over x_1^{\prime} + x_2^{\prime}} \, \chi_1 \right ) \,
\left [ 2 \, {\rm Li}_2 \left ( 1 - { x_2 \over x_2^{\prime}} \right )
+ \ln^2 { x_2 \over x_2^{\prime}}  - \ln^2 { x_2 \over x_1^{\prime}} \right ]
\nonumber \\
&& - \, {2 \, x_1  \over x_1^{\prime} \, (x_1^{\prime} + x_2^{\prime}) } \, \chi_1 \,
\left [ 2 \, {\rm Li}_2 \left (  { x_1 \over x_1^{\prime}} \right )
- {3 \over 2}  \, { x_1^{\prime} \over x_2^{\prime} }  \,
\ln{ x_1^{\prime} \over x_1^{\prime} + x_2^{\prime}}
- {3 \over 2} \, { x_2 \over x_1}  \, \ln{ x_2 \over x_1^{\prime} + x_2^{\prime}}
- { \pi^2 \over 3 } - 1 \right ]
\nonumber \\
&& - \, {2 \, x_1  \over x_1^{\prime} \, (x_1^{\prime} + x_2^{\prime}) } \,
\, \left (\chi_1 - \chi_4 \right ) \,
\left (  \ln { x_1 \over x_1^{\prime}}  + \ln { x_2 \over x_2^{\prime}} \right )
+ {16 \over x_1^{\prime}} \, \ln { x_1^{\prime} \over x_1^{\prime} - x_1} \,,
\\
\nonumber \\
\mathbb{V}^{(1), \, C_F^2}_{\rm 3P} &=&  \Bigg [ 2 \, \kappa_1 \,
{ \color{blue} \bigg ( } { \mathbb{G}_1  \over x_1^{\prime} \, x_3^{\prime}}  \, \theta(x_2 - x_2^{\prime}) \,
- {x_3 \, \mathbb{G}_2  \over x_1^{\prime} \, x_2^{\prime} \, \bar x_2 \, \bar x_2^{\prime} }
+ { \bar x_1^{\prime} \, x_3 \, \mathbb{G}_3
\over x_1^{\prime} \, x_2^{\prime} \, x_3^{\prime} \, (\bar x_1^{\prime} - x_2 ) }  \, \theta(x_1 - x_1^{\prime} )
+ { x_3 \, \, \mathbb{G}_4  \over x_3^{\prime} \, \bar x_2^{\prime} \, (x_2 - x_2^{\prime}) } \,
\theta(x_1 - \bar x_2^{\prime} )
\nonumber \\
&& \hspace{0.3 cm} - \, { \bar x_1 \, \mathbb{G}_5 \over x_2^{\prime} \, \bar x_2^{\prime} }
\, \theta(x_2 - x_2^{\prime})   { \color{blue} \bigg ) }
 + {1 \over 2} \, \chi_3 \, \mathbb{G}_6
+  {8  \, \mathbb{G}_7  \over x_3^{\prime} \, (x_2 - x_2^{\prime}) }
\, \theta(x_2 - x_2^{\prime} )
-  2 \, ( \kappa_1  -  \kappa_4)  \, \left ( { \mathbb{G}_8  \over x_2 - x_2^{\prime}}  + \mathbb{G}_9  \right  )
\nonumber \\
&&  -  \, \left ( {\eta_1  \, \kappa_4 - \eta_4  \, \kappa_1 \over 2 }
-   \eta_1 -   \kappa_1 +   \chi_1 +   { \chi_3 \over 2 } \right )  \,
\mathbb{G}_{10}    \Bigg ]_{\oplus}
 +  \,  \left [ x_1 \leftrightarrow x_2,  \, x_1^{\prime}  \leftrightarrow  x_2^{\prime},
\, \eta_i  \leftrightarrow \kappa_i \right ]\,,
\end{eqnarray}
where $\{ \eta_1, \, \eta_2, \, \eta_3,  \, \eta_4 \}=\{ 4, \, 16, \, 0,  \, 4 \}$,
$\{ \kappa_1, \, \kappa_2, \, \kappa_3,  \, \kappa_4 \}=\{ 0, \, 0, \, 16,  \, 0 \}$,
and $\{ \chi_1, \, \chi_2, \, \chi_3,  \, \chi_4 \}=\{ 4, \, 16, \, 0,  \, 8 \}$.
For brevity we have further defined  the  coefficient  functions $\mathbb{G}_{1,..., 10}$ to
describe  the genuine three-particle contribution
\begin{eqnarray}
\mathbb{G}_{1} &=&  { x_1 - x_1^{\prime} \over x_2 - x_2^{\prime}} \,
\left [  \left (\ln { x_3  \over \bar x_2} + \ln { x_1^{\prime} \over x_1^{\prime} - x_1}   \right )
\, \theta(x_2 - \bar x_1^{\prime} )
-  \left (  \ln { x_3  \over \bar x_1^{\prime} - x_2}
+   \ln { x_3^{\prime}  \over \bar x_2^{\prime} - x_1}   \right  )  \,  \theta(x_1 -  x_1^{\prime} )
- \ln {x_3  \over \bar x_2} -  \ln {\bar x_2^{\prime}  \over \bar x_2^{\prime} - x_1} \right ]
\nonumber \\
&& - \,  { x_1 \over \bar x_2} \, { x_2 - \bar  x_1^{\prime} \over x_2 -  x_2^{\prime} } \,
\left [  \left ( \ln { \bar x_2^{\prime} \over x_1^{\prime} }
+  \ln { x_2 - \bar  x_1^{\prime} \over x_2 -  x_2^{\prime} } \right )
\, \theta(x_2 - \bar x_1^{\prime})
+ \left ( \ln { x_1 \, x_3 \over \bar x_2^2}  + \ln { \bar x_2^{\prime}  \over x_2 - x_2^{\prime} }   \, \right )
\, \theta(\bar x_1^{\prime} - x_2)  \right ]
\nonumber \\
&& +  \,  { x_1  \over \bar x_2^{\prime}}  \, { x_3  \over x_2 -  x_2^{\prime} }
{x_3^{\prime} \over \bar x_2^{\prime}  - x_1} \,
\left  [ \ln  {x_3  \over x_1}  +  \ln { \bar x_2^{\prime} \over \bar x_2^{\prime} - x_1 }
+  \ln { x_2 -   x_2^{\prime} \over \bar x_2^{\prime} - x_1 } \right ]
\nonumber \\
&& +  \,  { x_1 -  x_1^{\prime} \over  x_1 -  \bar x_2^{\prime} }
\left [ \ln{ x_1 -  x_1^{\prime} \over x_2 -  x_2^{\prime} }
- \ln { x_3  x_3^{\prime} \over (\bar x_2^{\prime} - x_1)^2 }  \right ]   \theta(x_1 - x_1^{\prime}) \,,
\hspace{0.70 cm}  \\
\nonumber \\
\mathbb{G}_{2} &=&  \ln { x_2 \, \bar x_2^{\prime} \over x_2 -  x_2^{\prime}}  \, \theta(x_2 -  x_2^{\prime})
+ \left [ {x_1 \, x_2 \over x_3} \, \left ( \ln { x_1 \over x_2} + \ln {\bar  x_1 \over \bar x_2}  \right )
- \ln{ x_3 \over \bar x_1 \, \bar x_2} \right ]
\, \theta(x_2^{\prime} - x_2)  \,,
 \\
\nonumber \\
\mathbb{G}_{3} &=&  \left ( \ln { x_3^{\prime} \over \bar x_1^{\prime}}
+ \ln {x_2 \over x_2 -  x_2^{\prime}} \right ) \, \theta(x_2 - x_2^{\prime})
- \left [ \ln { \bar x_1^{\prime} \over \bar x_1}   + \ln{x_3  \over \bar x_1^{\prime} - x_2}
+ {x_2 \over x_3} \, { x_1 -  x_1^{\prime} \over \bar x_1^{\prime}}  \,
\left ( \ln { x_2 \over \bar x_1 }  + \ln { \bar x_1^{\prime} - x_2 \over x_1 - x_1^{\prime}} \right ) \right ]
\,  \theta(x_2^{\prime} - x_2)  \,,
\hspace{0.8 cm}  \\
\nonumber \\
\mathbb{G}_{4} &=&  {x_2 \over x_3} \, { x_1 - \bar x_2^{\prime} \over x_2^{\prime}} \,
\left (  \ln { \bar x_1 \over x_2} + \ln{ x_1 - \bar x_2^{\prime} \over  x_2^{\prime} -x_2}  \right )
- \left ( \ln { x_3 \over \bar x_1 }  + \ln{ x_2^{\prime} \over x_2^{\prime} - x_2}  \right )  \,,
\\
\nonumber \\
\mathbb{G}_{5} &=&  \left (  \ln {x_2 \over \bar x_1} + \ln { \bar x_2^{\prime} - x_1 \over x_2 -  x_2^{\prime}}   \right ) \,
\left [ {1 \over x_3^{\prime} } \, \theta( x_1 - x_1^{\prime})
- {1 \over x_1^{\prime}} \, \theta(x_1^{\prime} - x_1)  \right ] \,,
\\
\nonumber \\
\mathbb{G}_{6} &=& { x_3  \over \bar x_1 \, x_1^{\prime} \, x_2^{\prime} \, \bar x_2^{\prime}} \, \ln x_1
- { x_3 \, \bar x_1^{\prime} \over \bar x_1 \, x_1^{\prime} \, x_2^{\prime}  \,x_3^{\prime}} \,
\ln { x_1 - x_1^{\prime} \over \bar x_1^{\prime} } \,  \theta( x_1 - x_1^{\prime})
+ { \bar x_1^{\prime} - x_2  \over x_1^{\prime} \, x_2^{\prime} \, x_3^{\prime}}
\, \ln { x_1 - x_1^{\prime} \over \bar x_1^{\prime} - x_2  }
\,  \theta( x_1 - x_1^{\prime}) \,  \theta( x_2^{\prime} - x_2 )
\nonumber \\
&&
+ { x_2^{\prime} - x_2 \over x_2^{\prime} \, \bar x_2^{\prime}}
\, \ln{ x_2 -  x_2^{\prime} \over  \bar x_2^{\prime} -  x_1 }
\, \left \{  {1 \over x_3^{\prime}}  \, \left [ \theta( x_1 - x_1^{\prime}) \, \theta( x_2 - x_2^{\prime})
+ \theta( x_1 - \bar x_2^{\prime}) \right ]
+ { x_1 \over x_1^{\prime} \, ( \bar x_2^{\prime} - x_1) }
\, \theta(x_1^{\prime} - x_1) \, \theta(x_2 - x_2^{\prime}) \right \}
\nonumber \\
&& + \, { x_3 \over  x_1^{\prime} \,  x_2^{\prime} \, (\bar x_2^{\prime} - x_1) } \,
\left [  \left ( \ln {x_3 \over x_1 }  + \ln {\bar x_2^{\prime}  \over \bar x_2^{\prime} - x_1} \right )
\, \theta(x_1^{\prime} - x_1)
- \, \left (  \ln{ x_1 \over \bar x_2^{\prime}}  + \ln { x_3^{\prime} \over x_1 - x_1^{\prime}} \right )
\, \theta(x_1 - x_1^{\prime}) \right ] \, \theta(x_2 - x_2^{\prime})
\nonumber \\
&& + \,  { x_3 \over  x_3^{\prime} \, \bar x_1}   \ln{ \bar x_1 \over x_3}
\left [ { \bar x_1^{\prime} \over x_1^{\prime}  \, x_2^{\prime}  }   \theta( x_1^{\prime} - x_1 )
- {1 \over \bar x_2^{\prime}}  \theta( \bar x_2^{\prime} - x_1 )  \right ]
-  { x_3 \over \bar x_1 \, \bar x_2^{\prime} \, x_3^{\prime} }
\, \ln{ x_2^{\prime}  \over x_1 - \bar x_2^{\prime} }  \, \theta(x_1 -\bar x_2^{\prime} )
\nonumber \\
&& + { x_2 \over \bar x_1 \,  x_2^{\prime} \, x_3^{\prime} } \, \ln{ x_2 \over \bar x_1 }
 \left [ { x_1 -  x_1^{\prime} \over  x_1^{\prime}}  \,    \theta( x_1^{\prime} - x_1 )
+ {\bar x_2^{\prime}  - x_1 \over \bar x_2^{\prime}  } \,  \theta(\bar x_2^{\prime} - x_1)  \right ]
- { \bar x_2 \over  x_1^{\prime} \,  x_2^{\prime} \, \bar x_2^{\prime} } \, \ln{ x_1 \over \bar x_2 }
\,\, \theta(x_2^{\prime} - x_2) \,,
\\
\nonumber \\
\mathbb{G}_{7} &=& { \bar x_2 \over x_1^{\prime} } \,
\left [ \ln{  x_3^{\prime} \over \bar x_1^{\prime} - x_2}  \, \theta(x_1 - x_1^{\prime})
- \ln { x_1 \over \bar x_2 } \, \theta(\bar x_1^{\prime} - x_2 )  \right ]
+ { x_1 \over x_1^{\prime}} \, { \bar x_1^{\prime} - x_2 \over x_1 -  x_1^{\prime}}
\, \ln{ x_1^{\prime} - x_1 \over x_2 -  \bar x_1^{\prime}}  \, \theta(x_2 - \bar x_1^{\prime})
\nonumber \\
&& - \, { x_3 \over x_1 -  x_1^{\prime}} \, \left ( \ln{ x_1^{\prime} \over x_1}
+ \ln{ x_3  \over x_1^{\prime}  - x_1} \right )  \, \theta(x_2 - \bar x_1^{\prime})
+  { x_3 \over x_1^{\prime}} \, { \bar x_2^{\prime}  \over \bar x_2^{\prime}  - x_1}
\, \left [ \ln{ x_1 \over \bar x_2^{\prime} }
- \ln { x_3 \over \bar x_2^{\prime}  - x_1} \, \theta(x_1^{\prime}  - x_1) \right ]
\nonumber \\
&& + \, {x_1 \over x_1^{\prime}} \, { x_2 - x_2^{\prime} \over \bar x_2^{\prime} - x_1}
\, \left [ \ln{ x_2 -  x_2^{\prime}  \over \bar x_2^{\prime} - x_1 } \, \theta(x_1^{\prime}  - x_1)
- \ln{ x_3^{\prime}  \over x_1 - x_1^{\prime} } \, \theta(x_1 - x_1^{\prime} )  \right ] \,,
 \\
\nonumber \\
\mathbb{G}_{8} &=&  {x_2  \over x_2^{\prime}} \, { \bar x_2^{\prime} - x_1 \over \bar x_2^{\prime}  \,  x_3^{\prime}}
\, \theta( x_1 - x_1^{\prime} ) \, \theta(\bar x_2^{\prime}  - x_1) \, \theta(x_2^{\prime} - x_2)
+ { x_1 \over x_1^{\prime}} \, { \bar x_2 - x_1^{\prime}   \over \bar x_2 \, x_3^{\prime}}
\, \theta(x_1^{\prime}  - x_1)  \, \theta(x_2 - x_2^{\prime}) \, \theta(\bar x_1^{\prime} - x_2 )
\nonumber \\
&& + \, { x_2 \over \bar x_2} \, { x_3 \over x_2^{\prime} \, (x_2 - \bar x_1^{\prime} )}
\, \theta(x_1 - x_1^{\prime}) \, \theta(x_2^{\prime}  - x_2)
- { x_1 \over x_1^{\prime}}  \, { x_3 \over \bar x_2^{\prime} \, (\bar x_2^{\prime} - x_1)  }
\, \theta(x_1^{\prime} - x_1) \, \theta(x_2 -  x_2^{\prime}) \,,
 \\
\nonumber \\
\mathbb{G}_{9} &=&   - {1  \over \bar x_2 \, \bar x_2^{\prime}} \,
\left [  { x_1 \, x_2 \over x_1^{\prime} \, x_2^{\prime}}
\, \theta(x_1^{\prime} - x_1) \, \theta(x_2^{\prime} - x_2)
- { x_3 \over x_3^{\prime}}  \, \theta(x_1 - x_1^{\prime}) \, \theta(x_2 - x_2^{\prime}) \right ] \,,
 \\
\nonumber \\
\mathbb{G}_{10} &=&  {1 \over \bar x_1 \, \bar x_2^{\prime}}
\, \left [ {x_2  \over x_2^{\prime}}  \, {\bar x_2^{\prime} - x_1 \over x_3^{\prime}}
\, \theta(x_1 - x_1^{\prime}) \,  \theta(x_2^{\prime} - x_2)
+ { x_3 \over x_3^{\prime}} \, \theta(x_1 - x_1^{\prime}) \, \theta(x_2 - x_2^{\prime})
+ { x_1  \over x_1^{\prime}} \, { x_3 \over \bar x_2^{\prime} - x_1 }
\, \theta(x_1^{\prime} - x_1) \, \theta(x_2 - x_2^{\prime})  \right .
\nonumber \\
&& \left . + \, { x_1 \over  x_1^{\prime}} \, { x_2 \over  x_2^{\prime}}
\, \theta(x_1^{\prime} - x_1) \, \theta(x_2^{\prime} - x_2)   \right  ]
\, \theta(\bar x_2^{\prime} - x_1) \,.
\end{eqnarray}
Here  we  have employed  the  customary ``bar notation" $\bar x_i \equiv 1 - x_i$,
$\bar x_i^{\prime} \equiv 1 - x_i^{\prime}$ (with $i=1, 2, 3$)
as well as    the generalized ``$\oplus$"  distribution  in the three variables $x_{1, 2, 3}$
\begin{eqnarray}
\left [f(x_1, \, x_2, \,  x_3, \,  x_1^{\prime}, \, x_2^{\prime}, \, x_3^{\prime}) \right ]_{\oplus}
& \equiv & f(x_1, \, x_2, \,  x_3, \,  x_1^{\prime}, \, x_2^{\prime}, \, x_3^{\prime})
  - \, \delta(x_1 - x_1^{\prime}) \,  \delta(x_2 - x_2^{\prime}) \,  \int [{\cal D} y] \,\,
f(y_1, \, y_2, \,   y_3, \,  x_1^{\prime}, \, x_2^{\prime}, \, x_3^{\prime})  \,.
\hspace{0.6 cm}
\end{eqnarray}
To highlight the essential role of introducing evanescent operators
in the extraction of the renormalization kernel $\mathbb{H}$ for the light-ray baryonic operator,
we proceed to write down explicitly  the analytic  expression for  the subtraction term
$\mathbb{Z}_{1 2}^{(1, 1)} \otimes  \, \mathbb{Z}_{2 1}^{(1, 0)}$
entering the  master formula   (\ref{master formula for the RG kernel})
\begin{eqnarray}
\mathbb{Z}_{1 2}^{(1, 1)} \otimes  \, \mathbb{Z}_{2 1}^{(1, 0)}
&=& {1 \over 2} \, \left ( {4 \, C_F \over N_c - 1} \right )^2  \,
\bigg  \{   -   \, {1 \over x_2^{\prime} +  x_3^{\prime}} \,   { \color{blue} \bigg  [ }
\left ( { x_2 \over x_3^{\prime}}  \, \ln{ x_2^{\prime} \over x_2^{\prime} +  x_3^{\prime}}
+ { x_3 \over x_2^{\prime}} \, \ln{ x_3 \over x_2^{\prime} +  x_3^{\prime}}
+ { x_2 \over x_2^{\prime}} \right )  \,  \theta(x_2^{\prime} - x_2)
\nonumber \\
&& \hspace{2.5 cm} + \,  \left ( { x_3 \over x_2^{\prime}}  \, \ln{ x_3^{\prime} \over x_2^{\prime} +  x_3^{\prime}}
+ { x_2 \over x_3^{\prime}} \, \ln{ x_2 \over x_2^{\prime} +  x_3^{\prime}}
+ { x_3 \over x_3^{\prime}} \right )  \,  \theta(x_3^{\prime} - x_3) { \color{blue} \bigg ] }
\, \delta(x_1 - x_1^{\prime})
\nonumber \\
&& + \,    \mathbb{G}_{10}(x_1, x_3, x_2, x_1^{\prime}, x_3^{\prime}, x_2^{\prime})
+   \, \mathbb{G}_{10}(x_3, x_2, x_1, x_3^{\prime}, x_2^{\prime}, x_1^{\prime})
-   \, \mathbb{G}_{10}(x_2, x_3, x_1, x_2^{\prime}, x_3^{\prime}, x_1^{\prime})  \bigg  \} \,.
\end{eqnarray}

\subsection{Matching Relations Between  Distinct Renormalization Schemes}

We are now in a position to present  the desired  result  of the short-distance coefficient function
governing  the matching relation for the leading-twist nucleon distribution amplitude
between the EO scheme and the KM scheme
\begin{eqnarray}
\mathbb{K}_{N}(x_i,  x_i^{\prime}, \mu)
= \delta(x_1 - x_1^{\prime}) \,  \delta(x_2 - x_2^{\prime})
+ \sum_{m =1}^{\infty}  \,  \left ({\alpha_s(\mu) \over 4 \pi} \right )^{m}
\, \mathbb{K}_{N}^{(m)}(x_i,  x_i^{\prime})  \,.
\end{eqnarray}
The  perturbative kernel function at the one-loop order  $\mathbb{K}_{N}^{(1)}$
has actually been  determined in our previous work \cite{Huang:2024ugd}
\begin{eqnarray}
\mathbb{K}_{N}^{(1)}(x_i,  x_i^{\prime}) &=&  \left ( { C_F \over N_c - 1} \right )  \,
\bigg \{  \left [ {3 \over x_1^{\prime} + x_2^{\prime}} \,
\left (  {x_1 \over x_1^{\prime}} \, \theta(x_1^{\prime} - x_1)
 + {x_2 \over x_2^{\prime}} \, \theta(x_2^{\prime} - x_2) \right )  \right ]_{+}
\, \delta(x_3 - x_3^{\prime})
\nonumber \\
&& \hspace{1.8 cm} +  \left [ {1 \over x_2^{\prime} + x_3^{\prime}} \,
\left (  {x_2 \over x_2^{\prime}} \, \theta(x_2^{\prime} - x_2)
+   {x_3 \over x_3^{\prime}} \, \theta(x_3^{\prime} - x_3) \right ) \right ]_{+}
\, \delta(x_1 - x_1^{\prime})
\nonumber \\
&& \hspace{1.8 cm} - \left [ {1 \over x_1^{\prime} + x_3^{\prime}} \,
\left (  {x_1 \over x_1^{\prime}} \, \theta(x_1^{\prime} - x_1)
 + {x_3 \over x_3^{\prime}} \, \theta(x_3^{\prime} - x_3) \right )  \right ]_{+}
\, \delta(x_2 - x_2^{\prime})
\nonumber \\
&& \hspace{1.8 cm} + \, {3 \over 2} \, \delta(x_1 - x_1^{\prime}) \, \delta(x_2 - x_2^{\prime}) \bigg \}  \,.
\end{eqnarray}
The yielding expression for the newly computed two-loop kernel function  $\mathbb{K}_{N}^{(2)}$  can be explicitly written as
\begin{eqnarray}
\mathbb{K}_{N}^{(2)}(x_i,  x_i^{\prime}) &=&
 {1 \over 4} \, \left \{ \,  \left [ \mathbb{H}^{(1)}(x_i, \,  x_i^{\prime})
 \left .  \right |_{{ \color{magenta} \eta_i \to \eta_i^{\prime}, \,\,
 \kappa_i \to \kappa_i^{\prime}, \,\, \chi_i \to \chi_i^{\prime}}}
 \right ]
 -  \left [ \mathbb{H}^{(1)}(x_i, \,  x_i^{\prime})
 \left .  \right |_{{ \color{magenta} \eta_i \to 0, \,\, \kappa_i \to 0, \,\,  \chi_i \to 0}}   \right ]  \right \}
\nonumber  \\
&&  - \, \left ( { C_F \over N_c - 1} \right )^2  \,
\bigg \{ \bigg [ {7 \over 2} \, {1 \over x_1^{\prime} + x_2^{\prime}}
\, { \color{blue} \bigg ( } \,
\left (  { x_1 \over x_2^{\prime} } \,  \ln {x_1^{\prime}  \over x_1^{\prime} + x_2^{\prime} }
+ { x_2 \over x_1^{\prime} } \, \ln{ x_2 \over x_1^{\prime} + x_2^{\prime} } + {x_1 \over x_1^{\prime}}   \right )
\, \theta(x_1^{\prime} - x_1)
\nonumber \\
&& \hspace{2.5 cm} + \, \,
\left (  { x_2 \over x_1^{\prime} } \,  \ln {x_2^{\prime}  \over x_1^{\prime} + x_2^{\prime} }
+ { x_1 \over x_2^{\prime}}  \, \ln{ x_1 \over x_1^{\prime} + x_2^{\prime} } + {x_2 \over x_2^{\prime}}   \right )
\, \theta(x_2^{\prime} - x_2)  { \color{blue} \bigg ) } \,  \, \delta(x_3 -  x_3^{\prime}) \bigg ]
\nonumber \\
&& \hspace{2.5 cm}  + \, \left [ x_1 \leftrightarrow x_3,  \, x_1^{\prime}  \leftrightarrow  x_3^{\prime}  \right ]
-  {1 \over 7} \, \left [ x_2 \leftrightarrow x_3,  \, x_2^{\prime}  \leftrightarrow  x_3^{\prime}  \right ]
{ \color{magenta} + \Delta \mathbb{K}_{N}^{(2)}(x_i,  x_i^{\prime}) } \bigg \} \,,
\end{eqnarray}
where $\{ \eta_1^{\prime}, \, \eta_2^{\prime}, \, \eta_3^{\prime},  \, \eta_4^{\prime} \}=\{ -2, \, 20, \, -8,  \, 0 \}$,
$\{ \kappa_1^{\prime}, \, \kappa_2^{\prime}, \, \kappa_3^{\prime},  \, \kappa_4^{\prime} \}=\{ 2, \, -4, \, - 16,  \, 0 \}$,
$\{ \chi_1^{\prime}, \, \chi_2^{\prime}, \, \chi_3^{\prime},  \, \chi_4^{\prime} \}=\{ -6, \, -20, \, 24,  \, -4 \}$.
Moreover, we have  introduced  the invariant function  $\Delta \mathbb{K}_{N}^{(2)}$ for brevity
\begin{eqnarray}
\Delta \mathbb{K}_{N}^{(2)}(x_i,  x_i^{\prime})  &=&   2
\left \{  \left [  \mathbb{G}_6(x_1, x_2, x_3, x_1^{\prime}, x_2^{\prime}, x_3^{\prime})
- \mathbb{G}_6(x_2, x_1, x_3, x_2^{\prime}, x_1^{\prime}, x_3^{\prime})  \right ]
- \left [ x_1 \leftrightarrow x_3,  \, x_1^{\prime}  \leftrightarrow  x_3^{\prime}  \right ]
- \left [ x_2 \leftrightarrow x_3,  \, x_2^{\prime}  \leftrightarrow  x_3^{\prime}  \right ]   \right \}
\nonumber \\
&& + \, {1 \over 2} \, \left [  5 \, \mathbb{G}_{10}(x_1, x_2, x_3, x_1^{\prime}, x_2^{\prime}, x_3^{\prime})
+ 5 \, \mathbb{G}_{10}(x_2, x_1, x_3, x_2^{\prime}, x_1^{\prime}, x_3^{\prime})
- 5 \, \mathbb{G}_{10}(x_1, x_3, x_2, x_1^{\prime}, x_3^{\prime}, x_2^{\prime}) \right .
\nonumber \\
&& \hspace{0.75 cm} \left . + \, 7 \,  \mathbb{G}_{10}(x_3, x_1, x_2, x_3^{\prime}, x_1^{\prime}, x_2^{\prime})
+ 7 \, \mathbb{G}_{10}(x_2, x_3, x_1, x_2^{\prime}, x_3^{\prime}, x_1^{\prime})
- 9 \, \mathbb{G}_{10}(x_3, x_2, x_1, x_3^{\prime}, x_2^{\prime}, x_1^{\prime})   \right ] \,.
\end{eqnarray}
It is then straightforward to derive the conversion formula  of the evolution  kernel $\mathbb{H}$
for the twist-three nucleon distribution amplitude between the two distinct renormalization schemes under discussion
\begin{eqnarray}
\mathbb{H}^{\rm KM}(x_i, \, x_i^{\prime}, \, \mu) &=&  \int [{\cal D} y] \, \int [{\cal D} y^{\prime}] \,
\left [ \mathbb{K}_{N}(x_i, \, y_i, \, \mu) \right ] \, \mathbb{H}^{\rm EO}(y_i, \, y_i^{\prime}, \, \mu)
\, \left [ \mathbb{K}_{N}(y_i^{\prime}, \,  x_i^{\prime}, \, \mu) \right ]^{-1}
\nonumber \\
&& + \, \int [{\cal D} y] \, \left [ \mathbb{K}_{N}(x_i, \, y_i, \, \mu) \right ] \,
{d \over d \ln \mu} \, \left [ \mathbb{K}_{N}(y_i, \,  x_i^{\prime}, \, \mu) \right ]^{-1}  \,.
\end{eqnarray}

\subsection{Renormalization-Scale Dependence of the Nucleon Distribution Amplitude}

We display in this section  the  manifest  expressions of  the  essential ingredients
entering the NLL evolution matrix $\mathbb{U}^{\rm NLL} $ for the non-perturbative shape parameters
$\Psi_{M m}$ of the leading-twist nucleon distribution amplitude $\Phi_{N}$.
For completeness and for convenience of  future phenomenological explorations,
we  would like to first summarize  the analytic expressions  for  eigenfunctions
of the one-loop renormalization kernel $\mathbb{H}^{(0)}$ in QCD
\begin{eqnarray}
{\cal P}_{M m}(x_i) = 120 \,  {\cal N}_{M m}^{-1} \, \sum_{Q=0}^{M} \, \sum_{q=0}^{Q}  \,
\mathbb{V}_{M m, \, Q q} \,\,  \Omega_{Q q}(x_i)
\equiv  120 \,  {\cal N}_{M m}^{-1} \,  \left [  \mathbb{V}  \,\,  \Omega \right ]_{M m} \,,
\label{expression of 1-loop eigenfunctions}
\end{eqnarray}
where the  basis functions  $\Omega_{Q q}(x_i)$   on the right-handed side
are suitably constructed to reproduce  the intrinsic  conformal properties of $\Phi_N(x_i, \mu)$ \cite{Braun:1999te}
\begin{eqnarray}
\Omega_{Q q}(x_i) = (Q + q +4) \, (x_1 + x_3)^{q} \, P_{Q-q}^{(2q+3, \, 1)}(2 x_2 -1) \,
C_q^{3/2} \left ({ x_1 - x_3 \over x_1 + x_3} \right ) \,.
\end{eqnarray}
Here $ P_{k}^{(\alpha, \, \beta)}(x)$ and $C_{n}^{3/2}(x)$ stand for Jacobi and Gegenbauer polynomials, respectively.
The determined values for the normalization coefficients ${\cal N}_{M m}$ with the truncation $M = 3$  can then be given by
\begin{eqnarray}
{\cal N}_{0 0} &=& 120, \qquad {\cal N}_{1 0} = 5040,  \qquad {\cal N}_{1 1} = 1680,
\qquad {\cal N}_{2 0} = 756,  \qquad {\cal N}_{2 1} = 3780,  \qquad {\cal N}_{2 2} = 216,
\nonumber \\
{\cal N}_{3 0} &=& {49 \over 44} \, \left (  1 +  { 247 \over 7 \, \sqrt{4801}}  \right ),
\qquad {\cal N}_{3 1} = {1925 \over 192} \, \left (  1 -  { 11 \over 5 \, \sqrt{97}}  \right ),
\qquad  {\cal N}_{3 2} = {1925 \over 192} \, \left (  1 +  { 11 \over 5 \, \sqrt{97}}  \right ),
\nonumber \\
{\cal N}_{3 3} &=& {49 \over 44} \, \left (  1 -  { 247 \over 7 \, \sqrt{4801}}  \right ).
\label{results of the normalization coefficients}
\end{eqnarray}
Applying the same truncation for the conformal expansion of the nucleon distribution amplitude
enables us to further write down  the  explicit   form  of  the  transformation matrix $\mathbb{V}$
\renewcommand{\arraystretch}{1.5}
\begin{eqnarray}
\mathbb{V} = \left(
                        \begin{array}{cccccccccc}
                          {1 \over 4} & 0 & 0 & 0 & 0 & 0 & 0 & 0 & 0 & 0 \\
                          0 & 0 & {7 \over 6} & 0 & 0 & 0 & 0 & 0 & 0 & 0 \\
                          0 & -{7 \over 10} & 0 & 0 & 0 & 0 & 0 & 0 & 0 & 0 \\
                          0 & 0 & 0 & {21 \over 100} & 0 & {63 \over 200} & 0 & 0 & 0 & 0 \\
                          0 & 0 & 0 & 0 & -{3 \over 4}  & 0 & 0 & 0 & 0 & 0 \\
                          0 & 0 & 0 & -{9 \over 50} & 0 & {21 \over 200} & 0 & 0 & 0 & 0 \\
                          0 & 0 & 0 & 0 & 0 & 0 & 0 & {11 \over 24 \, \sqrt{4801}} & 0 & {7 \over 800} + {247 \over 800 \, \sqrt{4801}} \\
                          0 & 0 & 0 & 0 & 0 & 0 & {11 \over 40 \, \sqrt{97}} & 0 &  {154 \over 7275 + 165 \, \sqrt{97}} & 0 \\
                          0 & 0 & 0 & 0 & 0 & 0 & -{11 \over 40 \, \sqrt{97}} & 0 &  {154 \over 7275 - 165 \, \sqrt{97}} & 0 \\
                          0 & 0 & 0 & 0 & 0 & 0 & 0 & -{11 \over 24 \, \sqrt{4801}} & 0 & {7 \over 800} - {247 \over 800 \, \sqrt{4801}} \\
                        \end{array}
                      \right) \,.
\end{eqnarray}
\renewcommand{\arraystretch}{1.0}
Hereafter we  collect  the  tensor  components   $\mathbb{X}_{M m, \, Q q}$  (with $0 \leq m, \, q \leq M, \, Q \leq 3$)
in a matrix form $\mathbb{X}$ whose elements  $\mathbb{X}_{i j}$  correspond  to  $\mathbb{X}_{M m, \, Q q}$
with $(M \, m)$ and $(Q \, q)$ being the $i$- and $j$-th element in the list   $\{ 00, \, 10,  \, ...,  \, 33 \}$
as ordered in  (\ref{results of the normalization coefficients}).
We are eventually  prepared to compile the analytic  results for   the one- and two-loop
anomalous dimension matrices $\bbGamma^{(0), \, (1)}$
dictating  the  renormalization-scale  evolution of the considered shape parameters in the EO scheme
\begin{eqnarray}
\bbGamma^{(0)} &=& \left ( {2 \, C_F \over N_c - 1} \right ) \,
{\rm diag} \left (1, \, {13 \over 3}, \, 5,  \, {19 \over 3},   \, {23 \over 3},  \, 8, \,
{ 559 - \sqrt{4801}\over 60}, \,  { 115 - \sqrt{97}  \over 12},  \,  { 115 + \sqrt{97}  \over 12},  \,
{ 559 + \sqrt{4801}\over 60} \right ) \,,
\nonumber \\
\bbGamma^{(1)} &=& \left ( \mathbb{V}^{\rm T} \right )^{-1} \, \widehat{\bbGamma}^{(1)} \, \mathbb{V}^{\rm T}  \,,
\qquad
\widehat{\bbGamma}^{(1)}  = \left ( {C_F \over N_c - 1} \right ) \,
\left [  C_A \,\, \widehat{\bbGamma}^{(1), \,\,  C_F \, C_A}
+   \beta_0 \,\, \widehat{\bbGamma}^{(1), \,\,  C_F \, \beta_0}
+  \left ( {C_F \over N_c - 1} \right )  \,\, \widehat{\bbGamma}^{(1), \,\,  C_F^2}  \right ]  \,,
\end{eqnarray}
where the newly introduced  matrix kernels $\widehat{\bbGamma}^{(1), \,\,  n}$
(with $n=C_F \, C_A,  \, C_F  \, \beta_0, \, C_F^2$) can be  derived as follows
\renewcommand{\arraystretch}{1.5}
\begin{eqnarray}
\widehat{\bbGamma}^{(1), \,\, C_F \, C_A} &=& \left(
                        \begin{array}{cccccccccc}
                          {38 \over 3} & 0 & 0 & 0 & 0 & 0 & 0 & 0 & 0 & 0 \\
                          0 & {46 \over 3} & 0 & 0 & 0 & 0 & 0 & 0 & 0 & 0 \\
                          0 & 0 & {26 \over 9} & 0 & 0 & 0 & 0 & 0 & 0 & 0 \\
                          0 & 0 & 0 & {493 \over 45} & 0 & -{34 \over 15} & 0 & 0 & 0 & 0 \\
                          0 & 0 & 0 & 0 & {7 \over 9}  & 0 & 0 & 0 & 0 & 0 \\
                          0 & 0 & 0 & -{119 \over 60} & 0 & -{17 \over 30} & 0 & 0 & 0 & 0 \\
                          0 & 0 & 0 & 0 & 0 & 0 & {91 \over 15} &  0 &  -{6 \over 35}  & 0 \\
                          0 & 0 & 0 & 0 & 0 & 0 & 0 & -{347 \over 126} & 0 &  -{775 \over 378} \\
                          0 & 0 & 0 & 0 & 0 & 0 & -{14 \over 135} & 0 &  -{997 \over 180} & 0 \\
                          0 & 0 & 0 & 0 & 0 & 0 & 0 & -{93\over 35} & 0 & -{1301 \over 252}  \\
                        \end{array}
                      \right)   \,,
\\
\nonumber \\
\widehat{\bbGamma}^{(1), \,\, C_F \, \beta_0} &=& \left(
                        \begin{array}{cccccccccc}
                          {19 \over 3} & 0 & 0 & 0 & 0 & 0 & 0 & 0 & 0 & 0 \\
                          0 & 16 & -{3 \over 5} & 0 & 0 & 0 & 0 & 0 & 0 & 0 \\
                          0 & -{5 \over 9}  & {128 \over 9} & 0 & 0 & 0 & 0 & 0 & 0 & 0 \\
                          {8 \over 3}  & {5 \over 6} & 0 & {2063 \over 90} & -{7 \over 15} & -{22 \over 15} & 0 & 0 & 0 & 0 \\
                          0 & 0 & -{15 \over 14}  & -{1 \over 3}  & {407 \over 18}  & -{4 \over 7}  & 0 & 0 & 0 & 0 \\
                          2 & -{5 \over 8}  & 0 & -{53 \over 40} & -{7 \over 10}  & {619 \over 30} & 0 & 0 & 0 & 0 \\
                          -{11 \over 14} & {22 \over 7} & 0 &  {209 \over 175} & 0 & {22 \over 25} & {4241 \over 150}
                           & -{12 \over 35} & -{363 \over 175} & -{1 \over 7}  \\
                          0 & 0 & {11 \over 7} & 0 & -{11 \over 90} & 0 & -{2 \over 9} & {1297 \over 45} &  -{1 \over 2} & -{47 \over 27} \\
                          {121 \over 108} &  {209 \over 216} & 0 & -{7117 \over 5400} & 0 & -{1144 \over 675} & -{3563 \over 2700}
                           & -{7 \over 15} &  {25399 \over 900} & -{14 \over 27} \\
                          0 & 0 & {3729 \over 1400} & 0 & -{143 \over 1000} & 0 & -{3 \over 25} & -{567 \over 250} & -{18 \over 25}
                           &  {2593 \over 100}  \\
                        \end{array}
                      \right)   \,,
\\
\nonumber \\
\widehat{\bbGamma}^{(1), \,\, C_F^2} &=& \left(
                        \begin{array}{cccccccccc}
                          -34 & 0 & 0 & 0 & 0 & 0 & 0 & 0 & 0 & 0 \\
                         {56 \over 15} & -56 &  {2 \over 5} & 0 & 0 & 0 & 0 & 0 & 0 & 0 \\
                          0 & -{10 \over 27}  & -{776 \over 27} & 0 & 0 & 0 & 0 & 0 & 0 & 0 \\
                          -{236 \over 15}  & -{10 \over 3} & 0 & -{1739 \over 27} & -{28 \over 45} & {22 \over 9} & 0 & 0 & 0 & 0 \\
                          0 & 0 &  {71 \over 7}  &  {4 \over 9}  & -{1091 \over 27}  &  {16 \over 21}  & 0 & 0 & 0 & 0 \\
                          -{149 \over 15} &  {10 \over 3}  & 0 &  {145 \over 36} & -{14 \over 15}  & -{665 \over 18} & 0 & 0 & 0 & 0 \\
                          {264 \over 35} & -{242 \over 21} & 0 &  -{341 \over 63} & 0 & -{286 \over 35} & -{1726 \over 25}
                           & -{86 \over 175} & -{159 \over 350} & -{29 \over 210}  \\
                          0 & 0 & -{77 \over 12} & 0 &  {143 \over 180} & 0 &  {43 \over 135} & -{462733 \over 9450} &  -{139 \over 420}
                          &  {24119 \over 5670} \\
                          -{539 \over 60} &  -{5137 \over 1944} & 0 &  {6545 \over 648} & 0 &  {1309 \over 90} &  {48517 \over 24300}
                           & {139 \over 450} &  -{8447 \over 200} &  {347 \over 405} \\
                          0 & 0 & -{5863 \over 600} & 0 & -{1573 \over 1000} & 0 &  {29 \over 250} &  {79252 \over 13125}
                          & -{1041 \over 875}  &  -{2599013 \over 63000}  \\
                        \end{array}
                      \right)   \,.
\end{eqnarray}
\renewcommand{\arraystretch}{1.0}
Additionally, we employ  the three-loop RG evolution of the strong coupling constant $\alpha_s(\mu)$
in the $\overline{\rm MS}$ scheme and the required expansion coefficients of the QCD $\beta$-function are taken from \cite{Baikov:2016tgj,Herzog:2017ohr}
\begin{eqnarray}
\beta_0 &=&  {11 \over 3} \, C_A - {4 \over 3} \, T_F \, n_f \,,
\hspace{5.0 cm}
\beta_1 =  {34 \over 3} \, C_A^2 - {20 \over 3} \, C_A \, T_F \, n_f -  4 \, C_F \, T_F \, n_f \,,
\nonumber \\
\beta_2 &=&  {2857 \over 54} \, C_A^3 - {1415 \over 27} \, C_A^2 \, T_F \, n_f
- {205 \over 9} \, C_F \, C_A \, T_F \, n_f  + 2 \, C_F^2 \, T_F \, n_f
+ {44 \over 9} \, C_F \, T_F^2 \, n_f^2  +  {158 \over 27}  \, C_A \, T_F^2 \, n_f^2  \,,
\end{eqnarray}
where $C_F = (N_c^2 -1)/ (2 \, N_c)$ and $C_A=N_c$ represent the quadratic Casimir operators of
the fundamental and adjoint representations of ${\rm SU}(N_c)$
with the standard  normalization $T_F =1/2$,
and $n_f$ stands for the number of quark flavours.

Furthermore,  we  collect here the derived  results of the conversion matrices  for
transforming the normalization constant and  shape parameters
in the EO scheme into the alternative KM scheme  at the two-loop  accuracy
\begin{eqnarray}
\Psi_{M m}^{\rm KM}(\mu) = \sum_{Q=0}^{M} \, \sum_{q=0}^{Q}  \,
\left [ \left ( \mathbb{V}^{\rm T} \right )^{-1} \,  \mathbb{\widehat{R}}  \,\,  \mathbb{V}^{\rm T} \right ]_{M m, \, Q q}
\, \Psi_{Q q}^{\rm EO}(\mu)   \,,
\end{eqnarray}
where the matrix kernel $\mathbb{\widehat{R}}$ can be perturbatively expanded in terms of the strong coupling  constant 
\begin{eqnarray}
\mathbb{\widehat{R}}  &=&  \mathbb{I}
+   \left ( { \alpha_s(\mu) \over  4 \, \pi} \right)
\left (  {C_F  \over N_c -1} \right ) \,\,  \mathbb{\widehat{R}}^{(1)}
+ \left ( { \alpha_s(\mu) \over  4 \, \pi} \right)^{2} \,\left (  {C_F  \over N_c -1} \right )
\left [ C_A \,  \mathbb{\widehat{R}}^{(2), \,\,  C_F \, C_A}
+ \beta_0 \, \mathbb{\widehat{R}}^{(2), \,\,  C_F \, \beta_0}
+  \left (  {C_F  \over N_c -1} \right )  \, \mathbb{\widehat{R}}^{(2), \,\,  C_F ^2}   \right ]
\nonumber \\
&& + \, {\cal O}(\alpha_s^3)  \,.
\end{eqnarray}
The  one- and two-loop conversion  matrices $\mathbb{\widehat{R}}^{(1)}$ and $\mathbb{\widehat{R}}^{(2), \,\, n}$
(with $n=C_F \, C_A,  \, C_F  \, \beta_0, \, C_F^2$) take the following forms
\renewcommand{\arraystretch}{1.5}
\begin{eqnarray}
\mathbb{\widehat{R}}^{(1)} &=& \left(
                               \begin{array}{cccccccccc}
                                 {3 \over 2} & 0 & 0 & 0 & 0 & 0 & 0 & 0 & 0 & 0 \\
                                 0 & {1 \over 2} & {3 \over 10} & 0 & 0 & 0 & 0 & 0 & 0 & 0 \\
                                 0 & {5 \over 18} & {3 \over 2} & 0 & 0 & 0 & 0 & 0 & 0 & 0 \\
                                 0 & 0 & 0 & {1 \over 10} & {7 \over 30} & {4 \over 15} & 0 & 0 & 0 & 0 \\
                                 0 & 0 & 0 & {1 \over 6} & {5 \over 6} & {2 \over 7} & 0 & 0 & 0 & 0 \\
                                 0 & 0 & 0 & {7 \over 30} & {7 \over 20} & {79 \over 60} & 0 & 0 & 0 & 0 \\
                                 0 & 0 & 0 & 0 & 0 & 0 & -{1 \over 10} & {6 \over 35} & {9 \over 35} & {1 \over 14} \\
                                 0 & 0 & 0 & 0 & 0 & 0 & {1 \over 9} & {1 \over 2} & {1 \over 4} & {5 \over 18} \\
                                 0 & 0 & 0 & 0 & 0 & 0 & {7 \over 45} & {7 \over 30} & {17 \over 20} & {7 \over 27} \\
                                 0 & 0 & 0 & 0 & 0 & 0 & {3 \over 50} & {9 \over 25} & {9 \over 25} & {23 \over 20} \\
                               \end{array}
                             \right)
\,,
\\
\nonumber \\
\mathbb{\widehat{R}}^{(2), \,\, C_F \, C_A } &=& \left(
                                                   \begin{array}{cccccccccc}
                                                   {33 \over 8} & 0 & 0 & 0 & 0 & 0 & 0 & 0 & 0 & 0 \\
                                                     0 & -{17 \over 48} & {13 \over 8} & 0 & 0 & 0 & 0 & 0 & 0 & 0 \\
                                                     0 & {325 \over 216} & {73 \over 16} & 0 & 0 & 0 & 0 & 0 & 0 & 0 \\
                                                     0 & 0 & 0 & -{817 \over 720} & {119 \over 180} & {89 \over 60} & 0 & 0 & 0 & 0 \\
                                                     0 & 0 & 0 & {17 \over 36} & {331 \over 144} & {359 \over 252} & 0 & 0 & 0 & 0 \\
                                                     0 & 0 & 0 & {623 \over 480} & {2513 \over 1440} & {67 \over 20} & 0 & 0 & 0 & 0 \\
                                                     0 & 0 & 0 & 0 & 0 & 0 & -{123 \over 80} & {17 \over 40} & {813 \over 1120}
                                                       & {5 \over 9} \\
                                                     0 & 0 & 0 & 0 & 0 & 0 & {119 \over 432} & {241 \over 224} & {307 \over 336}
                                                       & {17375 \over 12096} \\
                                                     0 & 0 & 0 & 0 & 0 & 0 & {1897 \over 4320} & {307 \over 360} & {6583 \over 2880}
                                                       & {3035 \over 2592} \\
                                                     0 & 0 & 0 & 0 & 0 & 0 & {7 \over 15} & {417 \over 224} & {1821 \over 1120}
                                                       & {3629 \over 1344} \\
                                                   \end{array}
                                                 \right)
\,,
\\
\nonumber \\
\mathbb{\widehat{R}}^{(2), \,\, C_F \, \beta_0 } &=&  \left(
                                                        \begin{array}{cccccccccc}
                                                          {5 \over 8} & 0 & 0 & 0 & 0 & 0 & 0 & 0 & 0 & 0 \\
                                                          0 & {11 \over 24} & {1 \over 8} & 0 & 0 & 0 & 0 & 0 & 0 & 0 \\
                                                          0 & {25 \over 216} & {13 \over 24} & 0 & 0 & 0 & 0 & 0 & 0 & 0 \\
                                                          -{2 \over 15} & -{1 \over 24} & -{3 \over 80} & {19 \over 60} & {91 \over 720}
                                                            & {1 \over 90} & 0 & 0 & 0 & 0 \\
                                                          {1 \over 21} & {5 \over 336} & {3 \over 56} & {97 \over 1008} & {7 \over 18}
                                                            & {11 \over 84} & 0 & 0 & 0 & 0 \\
                                                          {1 \over 120} & -{7 \over 192} & -{3 \over 320} & {73 \over 2880} & {49 \over 320}
                                                            & {739 \over 1440} & 0 & 0 & 0 & 0 \\
                                                          {11 \over 280} & -{11 \over 112} & -{33 \over 1120} & -{121 \over 2800}
                                                            & -{33 \over 800} & -{11 \over 175} & {67 \over 300} & {151 \over 1400}
                                                            & {51 \over 1400} & {23 \over 1680} \\
                                                         {11 \over 2016} & {55 \over 4032} & -{11 \over 224} & {11 \over 6720}
                                                            & -{11 \over 1440} & -{11 \over 630} & {323 \over 4320} & {39 \over 140}
                                                            & {449 \over 3360} & {71 \over 3024} \\
                                                          -{11 \over 480} & {11 \over 192} & {11 \over 240} & -{451 \over 43200}
                                                            & {77 \over 2700} & {209 \over 2700} & {1001 \over 21600} & {119 \over 900}
                                                            & {303 \over 800} & {847 \over 6480} \\
                                                          {11 \over 1120} & -{11 \over 1120} & -{297 \over 11200} & -{187 \over 28000}
                                                            & -{341 \over 8000} & -{143 \over 7000} & {67 \over 4000} & {489 \over 14000}
                                                            & {2307 \over 14000} & {8053 \over 16800} \\
                                                        \end{array}
                                                      \right)
\,,
\\
\nonumber \\
\mathbb{\widehat{R}}^{(2), \,\, C_F^2 } &=& \left(
                                              \begin{array}{cccccccccc}
                                                -{143 \over 16} & 0 & 0 & 0 & 0 & 0 & 0 & 0 & 0 & 0 \\
                                                -{7 \over 5} & -{17 \over 144} & -{85 \over 24} & 0 & 0 & 0 & 0 & 0 & 0 & 0 \\
                                                -{35 \over 36} & -{1765 \over 648} & -{4307 \over 432} & 0 & 0 & 0 & 0 & 0 & 0 & 0 \\
                                               {26 \over 15} & -{31 \over 96} & {37 \over 40} & {1337 \over 720} & -{539 \over 360}
                                                  & -{6119 \over 2160} & 0 & 0 & 0 & 0 \\
                                                -{47 \over 84} & -{55 \over 168} & -{57 \over 32} & -{1525 \over 1512} & -{1169 \over 216}
                                                  & -{10 \over 3} & 0 & 0 & 0 & 0 \\
                                                {29 \over 240} & {19 \over 64} & -{37 \over 40} & -{40703 \over 17280} & -{7721 \over 2160}
                                                  & -{15007 \over 1920} & 0 & 0 & 0 & 0 \\
                                                -{1023 \over 1120} & {1639 \over 2016} & {143 \over 672} & -{6457 \over 50400}
                                                  & {7139 \over 7200} & {583 \over 900} & {9061 \over 3000} & -{5659 \over 5250}
                                                  & -{20317 \over 14000} & -{56663 \over 50400} \\
                                                -{33 \over 224} & -{4015 \over 36288} & {2629 \over 4032} & -{1507 \over 20160}
                                                  & -{3443 \over 8640} & {3091 \over 15120} & -{4417 \over 7200} & -{834923 \over 302400}
                                                  & -{208253 \over 100800} & -{500923 \over 181440} \\
                                                {869 \over 2160} & -{1771 \over 5184} & -{77 \over 288} & -{29557 \over 129600}
                                                  & -{20713 \over 32400} & -{63613 \over 32400} & -{76447 \over 81000}
                                                  & -{74899 \over 36000} & -{297763 \over 54000} & -{285241 \over 97200} \\
                                                -{1639 \over 5600} & {1133 \over 5040} & {5753 \over 11200} & -{3729 \over 28000}
                                                  & {649 \over 1600} & -{25201 \over 42000} & -{39571 \over 45000} & -{475513 \over 140000}
                                                  & -{242321 \over 70000} & -{185033 \over 28000} \\
                                              \end{array}
                                            \right)
 \,.
\end{eqnarray}
\renewcommand{\arraystretch}{1.0}

\end{widetext}

\bibliographystyle{apsrev4-1}

\bibliography{References}

\begin{thebibliography}{90}
\bibitem [{Err()}]{Erratum-rho-meson-RGE}%
  \BibitemOpen
  \href@noop {} {}\bibinfo {note} {We have idendified an obvious misprint in
  the obtained expression of the NLO renormalization kernel for the transverse
  distribution amplitude $\Phi_{\perp}$ \cite{Mikhailov:2008my}, where the
  double-logarithmic term {``$(1/2) \, \ln^2 x$"} in the second line of (3.13)
  should be replaced by {``$(1/2) \, \ln^2 y$"} instead.}\BibitemShut {Stop}%
\end{thebibliography}

\end{document}